\documentclass{emulateapj}
\usepackage{amsmath,color}
\usepackage{comment}
\usepackage{subfigure}
\usepackage[normalem]{ulem}

\newcommand{\vek}[1]{\boldsymbol{#1}}
\newcommand{\rate}{${\rm Gpc}^{-3}{\rm yr}^{-1}$}

\begin{document}

\title{Calibrating the cosmic distance ladder using gravitational-wave observations}

\author{Anuradha Gupta}
\email{axg645@psu.edu}
\affiliation{Institute for Gravitation and Cosmos, Physics Department, Pennsylvania State University, University Park, PA, 16802, USA}

\author{Derek Fox}
\affiliation{Institute for Gravitation and the Cosmos, Physics Department, Pennsylvania State University, University Park, PA, 16802, USA}
\affiliation{Department of Astronomy \& Astrophysics, Pennsylvania State University, University Park, PA, 16802, USA}
\email{dbf11@psu.edu}

\author{B. S. Sathyaprakash}
\email{bss25@psu.edu}
\affiliation{Institute for Gravitation and the Cosmos, Department of Physics, Pennsylvania State University, University Park, PA, 16802, USA}
\affiliation{Department of Astronomy \& Astrophysics, Pennsylvania State University, University Park, PA, 16802, USA}
\affiliation{School of Physics and Astronomy, Cardiff University, Cardiff, UK, CF24 3AA}

\author{B. F. Schutz}
\affiliation{School of Physics and Astronomy, Cardiff University, 5, The Parade, Cardiff, UK, CF24 3AA}
\affiliation{Max Planck Institute for Gravitational Physics (Albert Einstein Institute), 14476 Potsdam/Golm, Germany}
\email{SchutzBF@cardiff.ac.uk}

\begin{abstract} 
Type Ia supernovae (SNe Ia) are among preeminent distance ladders for precision cosmology due to their intrinsic brightness, which allows them to be observable at high redshifts. Their usefulness as unbiased estimators of absolute cosmological 
distances however depends on accurate understanding of their intrinsic 
brightness, or anchoring their distance scale.
This knowledge is based on calibrating their distances with Cepheids. Gravitational waves from compact binary coalescences, being standard sirens, can be used to validate distances to SNe Ia, when both occur in the same galaxy or galaxy cluster. The current measurement of distances by the advanced LIGO and Virgo detector network suffers from large statistical errors ($\sim 50\%$). However, we find that using a third generation gravitational-wave detector network, standard sirens will allow us to measure distances with an accuracy of $\sim 0.1\%$-$3\%$ for sources within $\le300$ Mpc. These are much smaller than the dominant systematic error of $\sim 5\%$ due to radial peculiar velocity of host galaxies. Therefore, gravitational-wave observations could soon add a new cosmic distance ladder for an independent calibration of distances to SNe Ia.
\end{abstract}


\section{Introduction} 
The geometry and dynamics of the universe can be inferred by two key ingredients obtained for a population of cosmological sources: precise measurement of their redshift and accurate estimation of their luminosity distance. The luminosity distance $D_L$ to a source at a redshift $z$ depends on a number of parameters such as the Hubble-Lema\^{i}tre parameter $H_0,$ dimensionless dark matter and dark energy densities $\Omega_M$ and $\Omega_\Lambda,$ dark energy equation of state parameter $w$ (which may itself depend on redshift), and the curvature of space $\Omega_k.$ One can fit a cosmological model $D_L(z; \vec p)$ to a set of, say $n$, measurements $\{D^{\ell}_L, z_{\ell}\},$ $\ell=1,\ldots,n$ and hence determine the parameters $\vec p = (H_0, \Omega_M, \Omega_\Lambda, \Omega_k, w).$  It is apparent that to do so one must obtain an unbiased measurement of the distances and redshifts at cosmological scales.

\subsection{Standard Candles}
Distances can be measured using a \textit{standard candle} --- a source
whose intrinsic luminosity is well constrained, so that its measured
flux can be used to infer its distance.  Calibration of distance to
astronomical sources typically uses a ``distance ladder'' of multiple
steps to get from nearby sources to those at cosmological distances.
For example, in the most precise recent approach, nearby Type Ia
supernovae (SNe Ia) are calibrated via the ``standard candle''
behavior of Cepheid variable stars \citep{Riess:2019cxk}.
The Leavitt Law enabling determination of Cepheid luminosities from
their periods is calibrated in the Milky Way galaxy via
Cepheid parallaxes \citep{Riess:2018byc}; in the Large Magellanic
Cloud via observations of detached eclipsing binary systems
\citep{Pietrzynski:2013gia}; and in the ``megamaser'' galaxy NGC~4258
which has a known geometric distance from radio observations
\citep{Humphreys:2013eja}.

\subsection{Gravitational Wave Standard Sirens}
Observation of gravitational waves (GWs) has opened up the possibility of accurately measuring distances on all scales independent of the cosmic distance ladder.  Indeed, binary black holes and binary neutron stars are now being used to infer both the absolute and apparent luminosity of the source: the rate at which the emitted wave's frequency chirps up as it sweeps through the sensitivity band of a detector gives the source's intrinsic luminosity and the measured wave's amplitude gives the source's apparent luminosity. Combining the two we can infer the source's luminosity distance. The frequency evolution of the wave is completely determined by general relativity: it depends on the source's masses and spins, which are also measured via the wave's amplitude and frequency evolution in a network of detectors. Apart from general relativity, no detailed modeling of the source is required in this measurement. 

The apparent luminosity of the source (basically the strain amplitude) depends not only on the luminosity distance but also the source's position on the sky and the orientation of the binary's orbit relative to the line of sight from the detector to the source. With a network of three or more detectors it is, in principle, possible to infer all the unknown parameters of the source. In practice, however, the source's inclination is difficult to measure, especially when the orbital plane is close to face-on or face-off relative to the detector. This causes the biggest uncertainty in the estimation of luminosity distance of the source. In Sec.~\ref{sec:gw_as_siren}, we briefly discuss various uncertainties in the measurement of the source's luminosity distance from their GW signal.

Gravitational wave observations should be able to calibrate {\em all}
the rungs of the cosmic distance ladder for {\em every} galaxy or
galaxy cluster that hosts a binary merger, and have potential to
deliver new insights into the physics of these rungs. For example, one
can ask if the $D_n$-- $\sigma$ relationship, one of the rungs of the
distance ladder, is metallicity-dependent. Moreover, are there
systematic variations due to the inclination of the galaxy that could
be resolved from GW observations? Among all the rungs of distance
ladder, currently SNe~Ia are the only ones that can estimate
extragalactic distances at very high redshifts ($z\sim2.26$,
\citet{Rodney:2015cwa}) and have immense importance in
characterizing the cosmic expansion at $z<1$
\citep{Betoule:2014frx, Scolnic:2017caz}.
Accurate measurement of relative event-to-event distances to
SNe~Ia can be achieved via their well-characterized multicolor light
curve shapes \citep{rpk96} or, near-equivalently, their peak
luminosity--decline rate correlation \citep{pls+99}. However, while
precise relative calibration suffices to characterize the recent
cosmic expansion history to high-precision, thanks to the linearity
of the Hubble relation at $z \lesssim 0.1$, SNe~Ia can only support a
Hubble constant measurement via independent distance measurements
that provide an absolute calibration for their peak brightness.

\subsection{Standard Sirens for Measuring $H_0$}
\citet{Schutz86} noted that the standard siren property of compact binaries could be used as an independent measure of $H_0$ (also see \citet{Krolak1987}). However, the redshift $z$ to a merger event is degenerate with the binary's total mass $M$ and it is only possible to infer the combination $(1+z)M$ from GW measurements alone\footnote{In the case of binary neutron stars, tidal effects allow the determination of the redshift of a merger event \citep{read,Messenger:2013fya} albeit measurement errors based on current methods are too large to be useful for cosmography.}.  Unfortunately, the sky position error-box containing a merger event typically contains thousands of galaxies \citep{Gehrels:2015uga, Nair:2018ign}. Assuming the merger came from any of the galaxies within the error-box would lead to multiple values of $H_0$ for a single merger.  With a large enough population of events one gets a distribution of measured values of $H_0$ which will peak at its true value.  This way of estimating $H_0$ is known as {\it statistical} method and it does not require GW events to have an electromagnetic counterpart.  Alternatively, if electromagnetic follow-up observations in the sky position error-box of a merger identify a counterpart then it would be possible to directly obtain source's redshift \citep{Dalal:2006qt} and hence directly infer the Hubble-Lema\^{i}tre parameter.  Either of these methods requires accurate knowledge of the sky position of the source, which could be obtained with a network of three or more GW detectors.

\subsection{Current Status of $H_0$ Estimate}
Cepheid-based calibration of the nearby sample of SNe~Ia enables the
use of their counterpart SNe~Ia on cosmological scales to measure the
Hubble constant \citep{Riess:2016jrr, Riess:2019cxk}. This approach
currently gives $H_0 = 74.03 \pm 1.42$ km s$^{-1}$ Mpc$^{-1}$
\citep{Riess:2019cxk}. Calibrating these same supernovae via a
largely independent distance ladder based on the ``tip of the red
giant branch'' (TRGB) approach yields $H_0 = 
69.8 \pm 0.8$ (stat) $\pm 1.7$ (sys) km s$^{-1}$ Mpc$^{-1}$ \citep{fmh+19}.

An alternative geometric approach to distance measurement in
the late universe, by the H0LiCOW team, uses gravitational lensing
time delays and careful modeling to derive a somewhat less precise
single-step measurement of the Hubble constant, $H_0 =
73.3^{+1.7}_{-1.8}$ km s$^{-1}$ Mpc$^{-1}$
\citep{2019arXiv190704869W}.

Both of these $H_0$ values are larger than those derived from the
Planck Collaboration's observations of the cosmic microwave background
(CMB), $H_0 = 67.4 \pm 0.5$ km s$^{-1}$ Mpc$^{-1}$
\citep{Aghanim:2018eyx}, and from the $z\lesssim 2$ measurements of
the Baryon Acoustic Oscillation (BAO) peak of the galaxy correlation
function, as calibrated against the physical scale of the CMB acoustic
peak. The Dark Energy Survey (DES), for example, recently reported
$H_0 = 67.77 \pm 1.30$ km s$^{-1}$ Mpc$^{-1}$
\citep{Macaulay:2018fxi}, while a joint analysis of several recent BAO
results by \citet{Addison:2017fdm} gives $H_0 = 66.98\pm1.18$ km
s$^{-1}$ Mpc$^{-1}$.  Thus, present $H_0$ estimates can be divided
into two categories: early universe and CMB-calibrated
estimates (CMB, BAO) which tend low, and late universe estimates (SNe
Ia, H0LiCOW) which tend high (with the recent TRGB estimate in
between). The difference between the two classes of
measurement potentially reflects new physics on cosmological scales
\citep{Riess:2019cxk,2019arXiv190704869W}, either at low redshift or
in the early universe \citep{Aylor:2019jkm}.

After the detection of GW170817 \citep{GW170817} and identifying its host galaxy NGC 
4993 as an optical counterpart, it became possible to independently estimate the 
value of $H_0$, and \citet{GW170817_H0} reported it to be $70^{+12}_{-8}$ km s$^{-1}$ Mpc$^{-1}$.
As a proof-of-principle demonstration of the statistical method, $H_0$ was found to be 
$H_0 = 77^{+37}_{-18}$ km s$^{-1}$ Mpc$^{-1}$ without using the knowledge of 
NGC 4993 but the distance information from GW170817 alone \citep{Fishbach:2018gjp}. 
 \citet{Hotokezaka:2018dfi} reported an improved measurement of the Hubble 
constant of $70^{+5.3}_{-5.0}\,\rm km\,s^{-1}\,Mpc^{-1}$ when including an estimate of 
the inclination angle of the binary determined from radio observations of GW170817 \citep{2018Natur.561..355M}. 
\citet{Abbott:2019yzh} deployed the statical method on the population of binary black holes 
detected during the first and second observing runs of advanced LIGO and advanced Virgo detectors \citep{LIGOScientific:2018mvr} 
to obtain a value of $H_0 = 68^{+14}_{-7}\,\rm km\,s^{-1}\,Mpc^{-1}.$

\subsection{Calibrating SNe~Ia in Nearby Clusters with Standard Sirens}
SNe~Ia are believed to be the result of accretion induced collapse and
explosion of white dwarfs.  It is likely, however, that some of the
SNe~Ia come from mergers of binary white dwarfs instead of collapse of
accreting white dwarfs \citep{Raskin2012}. Distinguishing between
different subclasses of SNe~Ia, or between properties of
SNe~Ia discovered in early-versus late-type galaxies
\citep{jrs+18}, could be one of the applications of standard
sirens.

If SNe~Ia and binary neutron star mergers occur in the same galaxy or
galaxy cluster, it is possible to directly calibrate SNe~Ia
luminosities with distances inferred from GW observations. It is this
approach that we focus on in the present work. While it is highly
unlikely for a binary neutron star merger to occur in the same galaxy
as a SN~Ia in a given year, every merger event in a rich galaxy
cluster will typically be accompanied by multiple SNe~Ia from the
galaxies in that cluster. Considering only clusters rich enough to
host on average one or more SNe~Ia per year, we expect $\sim3.8$ SNe Ia per  
binary neutron star merger host galaxy per year of optical observation 
from the nearest 34 such clusters \citep{Girardi:2002mmg}, located at redshifts $z<0.072$
($D_L \lesssim 300$\,Mpc).  
Thus, GW observations from binary neutron star mergers provide a
unique opportunity to calibrate SNe~Ia and to look for subclasses of
SNe~Ia, which could improve the precision of using them as standard
candles.

Consistency of the Hubble diagram determined from GW and SNe Ia would confirm that calibration of SNe Ia is unlikely to have any systematic errors. On the contrary, any discrepancy in the Hubble flow determined by the two methods could point to systematics in either. One could, in principle, use the Hubble-Lema\^{i}tre parameter as a proxy for distance to SNe Ia hosts and calibrate their luminosities.  Such a calibration would work well on average but would not be useful for any one galaxy or galaxy cluster, as there are radial velocity departures from the Hubble flow that are unknown. Thus, it is necessary to know the peculiar velocity of the galaxy to infer the luminosity distance from $H_0$. However, if standard sirens and SNe Ia are both present in the same galaxy or galaxy cluster, the knowledge of the radial velocity is not needed for calibrating SNe Ia.

We note that the idea of calibrating SNe Ia using GWs distances has also been recently
explored by other authors \citep{Zhao:2017imr,Keeley:2019hmw}. 
Using only one binary neutron star merger GW170817, \citet{Zhao:2017imr} showed that the calibrations with 
both GWs and Cepheids lead to comparable SNe Ia light curves. \citet{Keeley:2019hmw}, on the other hand, emphasized 
on the importance of combining GW and SNe Ia data sets to achieve $1\%$ accuracy in the measurement of $H_0.$ 
The current paper is built on a similar idea and shows that it is possible to calibrate local SNe Ia distances 
with binary neutron star mergers occurring in the same galaxy cluster within $1\%$ accuracy using 3G  
GW detectors (it is almost impossible to find SNe Ia and binary neutron star mergers in the same galaxy though).

\section{Systematic Biases in the Measurement of Distance with Standard Sirens}
\label{sec:gw_as_siren}

In this section, we discuss various sources of systematic bias that can affect the distance measurement of GW sources.

\paragraph{Distance-Inclination Degeneracy}
The measurement of the luminosity distance $D_L$ is strongly correlated with that of the inclination angle $\iota$ of the binary with respect to the line-of-sight \citep{Ajith:2009fz,Usman:2018imj}. This is because both distance and inclination, along with the sky position angles, appear together in the amplitude of the GW polarization states (see, e.g., Eqs.~(2) in \citet{ACST94}). Due to this degeneracy, a face-on ($\iota=0^{\circ}$) or a face-off ($\iota=180^{\circ}$) binary far away has a similar GW amplitude to a closer edge-on ($\iota=90^{\circ}$) binary. This degeneracy can be broken to some extent by using a network having as many detectors as possible, as far away from each other on Earth as possible \citep{Cavalier:2006rz, Blair2008, Fairhurst2010, Wen:2010cr}. Employing accurate waveform models that incorporate higher harmonics and spin-precession also help break this degeneracy \citep{AMVISS09, TMPA2014, Vitale:2018wlg}. 
Measuring the event electromagnetically, if the binary coalescence has an electromagnetic counterpart,  partially
breaks the $D_L-\iota$ degeneracy \citep{Nissanke:2009kt}.
Moreover, if one can constrain the orbital inclination from the electromagnetic observations \citep{Evans:2017mmy}, the uncertainty in the distance measurement is greatly reduced as we will see below.  

\paragraph{Effect of Weak Lensing}
Gravitational waves just like electromagnetic waves get lensed when they propagate through the intervening matter \citep{Ohanian1974, Bliokh1975, Bontz1981, Deguchi1986, Nakamura1998}. The dark matter distribution along the line of sight as a GW propagates from its source to the detector can amplify or de-amplify signal's amplitude without affecting its frequency profile \citep{Wang:1996as, Dai:2016igl, Hannuksela:2019kle}. This `weak lensing' results in an additional random error in the distance measurement using GWs \citep{VanDenBroeck:2010fp}. \cite{KFHM06} showed that, in the case of super-massive black hole binaries, distance measurement error due to weak lensing dominates over other uncertainties leading to $\sim 6\%$ error for sources at $z=2$. This translates to $\sim0.1\%$ error for sources in the local universe ($\lesssim300$ Mpc) considered in this paper. We shall see below that this is less than the average error measured by a network of third generation GW detectors. Though there are proposals to remove the weak lensing effects substantially by mapping the mass distribution along the line of sight \citep{Gunnarsson:2005qu, Shapiro2010}, degradation of parameter estimation accuracy due to weak lensing will remain an issue for some time.  

\paragraph{Detector Calibration Errors}
It is important to note that the distance measurement is also affected by the detector calibration errors \citep{Abbott:2016jsd}. The uncertainty in the detector calibration implies an error in the measured amplitude and phase of the signal as a function of frequency. At present, the calibration error is between $5\%-10\%$ in amplitude and $3^{\circ}-10^{\circ}$ in phase over a frequency range of $20-2048$ Hz \citep{GW150914, GW151226, GW170104, GW170608, GW170814, GW170817}. As we will see in Sec.~\ref{sec:distance_measurement}, the median uncertainty in the measurement of distance to neutron star binary coalescences located at distances $\sim 10-300$\,Mpc  is $\sim 0.1\% - 3\%$, significantly smaller than the current calibration uncertainty in the amplitude. In addition to statistical errors, detector calibration may also suffer from small systematic errors. While these errors are expected to be small, there is currently no estimate of how large they might be. There is ongoing effort to improve the calibration of LIGO and Virgo detectors using alternative methods and it is expected that calibration errors will be sufficiently small to not significantly affect distance measurements \citep{Acernese:2018bfl,Abbott:2016jsd,Tuyenbayev:2016xey,Viets:2017yvy,Karki:2016pht}. These alternative methods should also help in understanding the systematic errors. 

In summary, GWs are `one-step' standard sirens (i.e., they do not require a calibrator at any distance), and hence, can provide unambiguous measurement of distance to the host galaxies and galaxy clusters in the local universe. This implies that GWs can be used as a distance indicator to calibrate nearby SNe Ia occurring in the same galaxy or galaxy cluster as the binary merger.

In the next section we investigate how probable is it to have binary merger and SNe Ia events in the same galaxy or galaxy cluster.

\begin{table*}[ht!]
\caption{Description of various detector networks used in this paper.}
\label{tab_det_net_info}
\begin{center}
\begin{tabular}{| c | c | c| c|} 
\hline
Network  & Detector location  & Detector sensitivity & $f_{\rm low}$ (Hz)\\  
\hline\hline
2G & Hanford-USA, Livingston-USA, Italy, India, Japan & aLIGO, aLIGO, AdV, aLIGO, KAGRA & 10, 10, 10, 10, 1 \\ 
\hline
3G & Utah-USA, Australia, Italy  & CE, CE, ET & 5, 5, 1\\
\hline
Hetero & Utah-USA, Livingston-USA, Italy, India, Japan & CE, Voyager, ET, Voyager, Voyager & 5, 5, 1, 5, 5\\
\hline
\end{tabular}
\end{center}
\end{table*}

\section{Spatial coincident observation of a binary neutron star merger and a Type Ia Supernova event}
\label{sec:rates}
Gravitational waves from a binary neutron star merger in the same galaxy as a SNe Ia
 could help calibrate the light curve of the latter and hence allow us to infer the 
luminosity function of SNe Ia. How likely is it to observe a binary 
coalescence in the same galaxy or galaxy cluster as a SNe Ia event? 

The current estimates of the local ($z=0$) SNe~Ia rate are in the range
$[2.38,\, 3.62] \times 10^4$ \rate\ with a median of $3.0 \times
10^4$ \rate\, \citep{2011MNRAS.412.1473L}, while that of binary neutron star mergers
are $[110,\, 3840]$ \rate\ with a median of
$\sim$1000 \rate~\citep{LIGOScientific:2018mvr}.  Using the SDSS
$r'$-band luminosity function of \citet{Blanton:2003hbb}, the number
density of galaxies in the local universe is $\approx10^7$
Gpc$^{-3}$, when integrated down to LMC-type ($0.1 L^*$) galaxies.
Hence, SNe~Ia occur at roughly once every 300 years per galaxy and
binary neutron star coalescences occur at a rate $\sim30$ times
smaller. Therefore, the chance of observing both of these events in a
single galaxy, over a ten year period, is roughly 1 in $10^3$ per
galaxy.

However, for every binary neutron star merger in a galaxy cluster one
expects to find a number of recent SNe~Ia.  Although the binary
neutron star merger rate in rich galaxy clusters is yet to be
measured, we assume it will track the SNe~Ia rate, as both populations
originate in compact object mergers. Hence, we anticipate the ratio of
SNe Ia and binary Neutron star merger volumetric rates 
$R_{\rm SN Ia}:R_{\rm BNS} \sim 30:1$ (estimated 90\%-confidence range of 8:1
to 300:1) will carry over to rich clusters directly.  Given an SNe~Ia
rate in $z<0.04$ rich galaxy clusters of 
$R_{\rm SN Ia} \sim [0.9, 1.4]\times 10^{-12}\,L_{B,\odot}^{-1}$ yr$^{-1}$, 
with a median of $1.2\times 10^{-12}\,L_{B,\odot}^{-1}$ yr$^{-1}$ \citep{Dilday2010},
this implies that there will be $\approx6$ SNe~Ia and $\sim0.2$ binary
neutron star mergers per year in a Coma-like cluster of total
luminosity $L_B\approx 5.0\times 10^{12}\,L_{B,\odot}$
\citep{Girardi:2002mmg}\footnote{Cluster SN~Ia rates at $z<0.5$
in these ``SNuB'' units from previous surveys (for $h=0.7$; 
\citealt{Dilday2010}) are: 1.16 \citep{mms+08}, 1.49
\citep{sgm+07}, 1.63 \citep{gms02}, and 1.29 \citep{gps+08}.}.

\begin{figure}[h!]\centering
\includegraphics[scale=0.5]{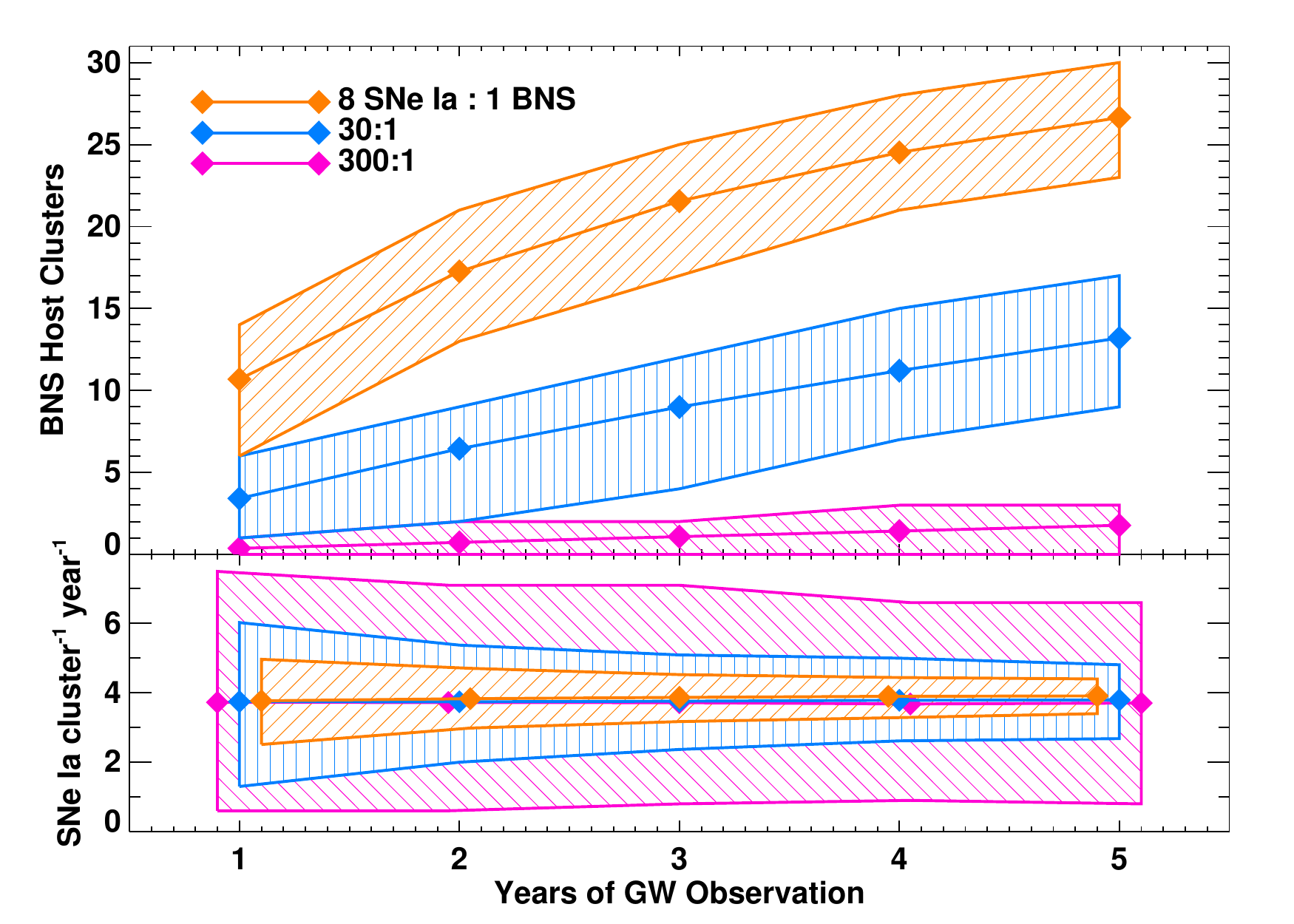}
\caption{Projected number of rich galaxy clusters with distances
calibrated by GW observation of binary neutron star mergers (BNS), as
a function of the ratio of rates of SNe~Ia to BNS mergers (8:1 in
orange; 30:1 in light blue; 300:1 in magenta) and duration of active
GW observations with appropriate sensitivity ($D_L\le
300\,(h/0.72)^{-1}$\,Mpc). Illustrated ranges are at
90\%-confidence. Upper panel: Number of rich galaxy clusters at
$z<0.072$ (out of 34 in the sample) which will host BNS mergers. Lower
panel: Rates of detection for SNe~Ia in the BNS host clusters, quoted
as rates per cluster per year of optical observations. Plot $x$
positions have been adjusted for clarity; all simulations were
evaluated at integer years only. See text for discussion.}
\label{fig:clsim}
\end{figure}

In order to explore the implications of binary distance measurements
for calibration of SNe~Ia luminosities, we consider a catalog of the
34 nearest ($z<0.072$; $D_L\le 300\,(h/0.72)^{-1}$\,Mpc) galaxy
clusters having luminosities $L_B \gtrsim 8\times 10^{11}\,
L_{B,\odot}$, sufficiently rich that each is expected to host one or
more SNe~Ia per year. Drawing cluster identifications and luminosities
from \citet{Girardi:2002mmg}, with redshifts from the NASA/IPAC
Extragalactic Database\footnote{NASA/IPAC Extragalactic
Database: \url{https://ned.ipac.caltech.edu}}, we carry out numerical
simulations of the number of binary neutron star mergers observed in
each cluster for active GW observing campaigns of duration 1~year to
5~years. Each simulation assumes a ratio of SNe~Ia to binary merger rates
of either 30:1 (median), 300:1 (pessimistic), or 8:1 (optimistic),
spanning the current 90\%-confidence range in binary neutron star merger
rates. Uncertainties in this ratio dominate over the present
uncertainty in the SNe~Ia rate for rich clusters.

These simulations seek to answer two questions: (1) how many cluster
distances can be calibrated by GW observation of binary neutron star
mergers; and (2) how many SNe~Ia luminosities can be calibrated, in
turn, via these cluster distances. Results are presented in
Fig.~\ref{fig:clsim}: The mean number of clusters with binary-neutron-star-merger-based
(GW-derived) distance measurements after 5~years of GW observation is
1.8, 13.2, and 26.6 clusters (of 34 in the sample) for the
pessimistic, median, and optimistic cases, respectively. The
90\%-confidence ranges on these estimates are roughly $\pm$4 in the
median and optimistic cases, and $\pm$1 in the pessimistic case. In
the pessimistic case, it is possible (with $\approx$1.6\% probability)
that we do not observe any cluster that hosts any binary neutron star event
 even after 5~years of GW
observation.

The number of SNe~Ia that can be calibrated via these binary merger
host clusters depends on the total duration and efficiency of
any associated optical observing campaign(s) capable of
discovering and characterizing SNe in these clusters. We therefore
estimate the rate of calibrated SNe~Ia per cluster per year of optical
observation, a metric that is relatively robust both to the ratio of
SNe~Ia to binary neutron star merger rates (whether optimistic,
median, or pessimistic), and to the duration of the GW observing
campaign. To estimate the total number of calibrated SNe~Ia, one
multiplies the per cluster per year rate (lower panel) by the number
of merger host clusters for the given GW year scenario (upper panel),
and by the duration of optical observations in years. As an
important caveat, we note that only SNe~Ia with high signal-to-noise
detections and either spectroscopy or high-quality multiband
photometry (or both) will likely be useful for precise absolute
calibration and Hubble constant measurement.

The main survey of the Large Synoptic Survey Telescope
\citep{lsst_2017} is planned to extend for ten years, and this
facility will be capable of discovering and characterizing the
majority of SNe~Ia in most of these clusters; as a caveat, we
note that not all considered clusters lie within the LSST survey
area, and clusters within the survey area will still be subject to
seasonal observability constraints. Similar considerations will
apply when estimating the useful yield from other optical surveys
seeking to characterize SNe~Ia in these galaxy clusters.
Overall, we consider a ten year period of observation
to be reasonable for the 2030's time frame of the GW
campaigns. As seen in Fig.~\ref{fig:clsim}, such a ten year baseline
of optical observations potentially enables calibration of
$\approx38$ SNe~Ia per binary neutron star merger host cluster. In the
upper panel, the number of unique binary neutron star host clusters does not
increase linearly with time since we have a finite number of
clusters and mergers repeatedly occur in some of the clusters; we note
that multiple binary neutron star mergers in the same cluster would
further improve the statistical uncertainty in the calibration
of that cluster's supernovae. 

We note that 90\%-confidence ranges on these numbers are larger than
the Poisson error on the number of SNe~Ia would suggest, because
fluctuations in the number of binary neutron star host clusters with
GW distance measurements typically dominates the overall
uncertainty. Overall, as a robust lower bound, Fig.~\ref{fig:clsim}
shows that the binary merger approach can anticipate successful
calibration of $>$1 SNe~Ia per cluster per year of high-quality
optical survey coverage, or $>$10 SNe~Ia per cluster for ten years
of optical observation.

In the next section, we compute the error in the measurement of distance to 
the nearby galaxy clusters hosting binary neutron star mergers and see how 
accurately we can estimate distances using various future networks of GW detectors.

\section{Distance Measurement Accuracy Using Standard Sirens}
\label{sec:distance_measurement}

Let us consider a population of binary neutron stars is uniformly distributed in the co-moving volume between luminosity distance $D_L$ of $10$ Mpc and $300$ Mpc. As we shall see below, for binary neutron star mergers closer than about 300 Mpc the statistical error in the distance measurement is well below systematic errors. Moreover, at such distances we can approximate the luminosity distance-redshift relation to be given by the Hubble-Lema\^{i}tre law $D_L = cz/H_0$ and we don't need to worry about cosmological effects. Also, since we will be using GWs to calibrate distance to SNe in the local universe, this distance range is more relevant. 

We assume neutron stars in the binaries to be non-spinning, have fixed masses $m_1=1.45M_{\odot}$ and  $m_2=1.35M_{\odot}$ and be located randomly on the sky; that is, their declination $\theta$ and right ascension $\phi$ obey uniform in $[-1,1]$ in $\sin\theta$ and uniform in $[0^{\circ}, 360^{\circ}]$ in $\phi,$ respectively. Further, we assume that the cosine of the inclination angle $\iota$ (the angle between binary's orbital angular momentum ${\bf L}$ and the line of sight ${\bf N}$) is uniform in $[-1, 1]$. The antenna pattern functions of GW detector  also depend on the polarization angle $\psi$, which sets the inclination of the  component of ${\bf L}$ orthogonal to ${\bf N}$ (see Sec.~4.2.1 in \citet{SathyaSchutzLivRev09}). We choose $\psi$ to be uniform in  $[0^{\circ}, 360^{\circ}]$. This constitutes the parameter space, $\{m_1, m_2, D_L, \iota, \theta, \phi, \psi, t_c, \phi_c \}$, for our target binary neutron stars, where $t_c$ and $\phi_c$ are the time and phase at the coalescence of the binary and we set them to be zero in our calculations.  
As binary neutron stars have long inspirals, we use 3.5PN accurate TaylorF2 waveform \citep{BIOPS2009} to model their GWs.

\begin{figure}[h!]\centering
\includegraphics[scale=0.55]{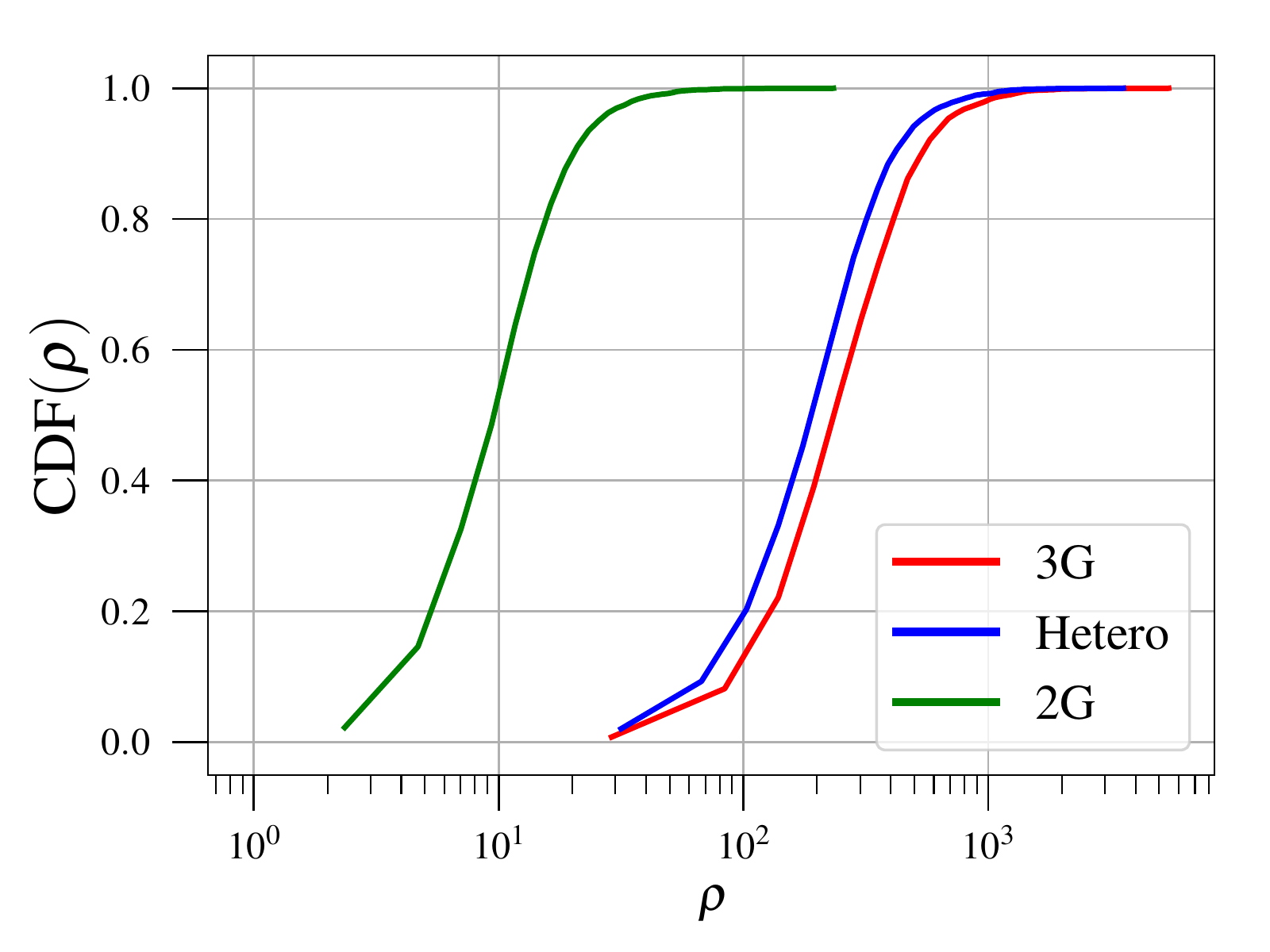}
\caption{Cumulative distribution of network SNR for 2G, 3G, Hetero networks, summarized in Tab.~\ref{tab_det_net_info}. 
A population of binary neutron stars with fixed masses $m_1 = 1.45M_{\odot}$ and $m_2 = 1.35M_{\odot}$ have isotropic sky-locations and orbital inclinations and 
are uniformly distributed in the co-moving volume between 10 Mpc and 300 Mpc.}
\label{fig:snr}
\end{figure}

\begin{figure*}[h!]\centering
\includegraphics[scale=0.48]{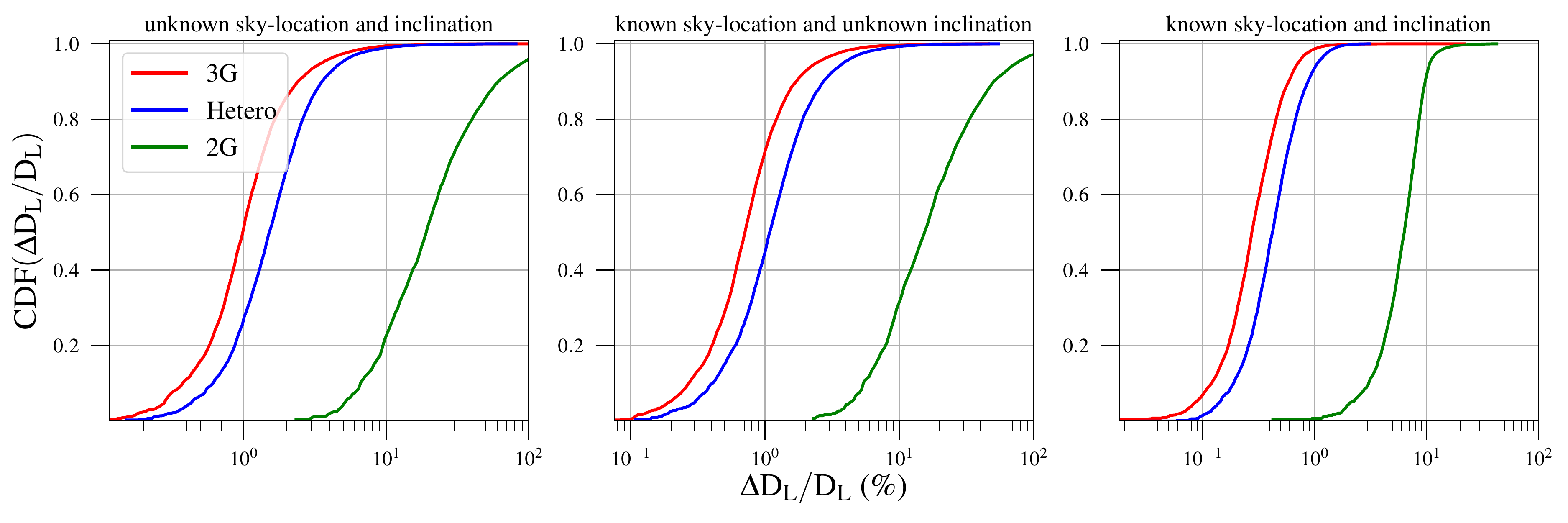}
\caption{Cumulative distribution of $1-\sigma$ distance errors measured with various networks of detectors, 2G, 3G, Hetero, summarized in Tab.~\ref{tab_det_net_info}. 
A population of binary neutron stars with fixed masses $m_1 = 1.45M_{\odot}$ and $m_2 = 1.35M_{\odot}$ have isotropic sky-locations and orbital inclinations and 
are uniformly distributed in the co-moving volume between 10 Mpc and 300 Mpc.
Left panel shows the errors when sky-location and orbital inclination of the binaries are not known to us. Middle panel shows the 
error when sky-location of the binaries are known and the right panel demonstrates distance errors when both sky-location and orbital inclination of binaries are 
known. All the sources plotted here have network SNR $\geq10$.}
\label{fig:dist}
\end{figure*}
\begin{figure*}[h!]\centering
\includegraphics[scale=0.65]{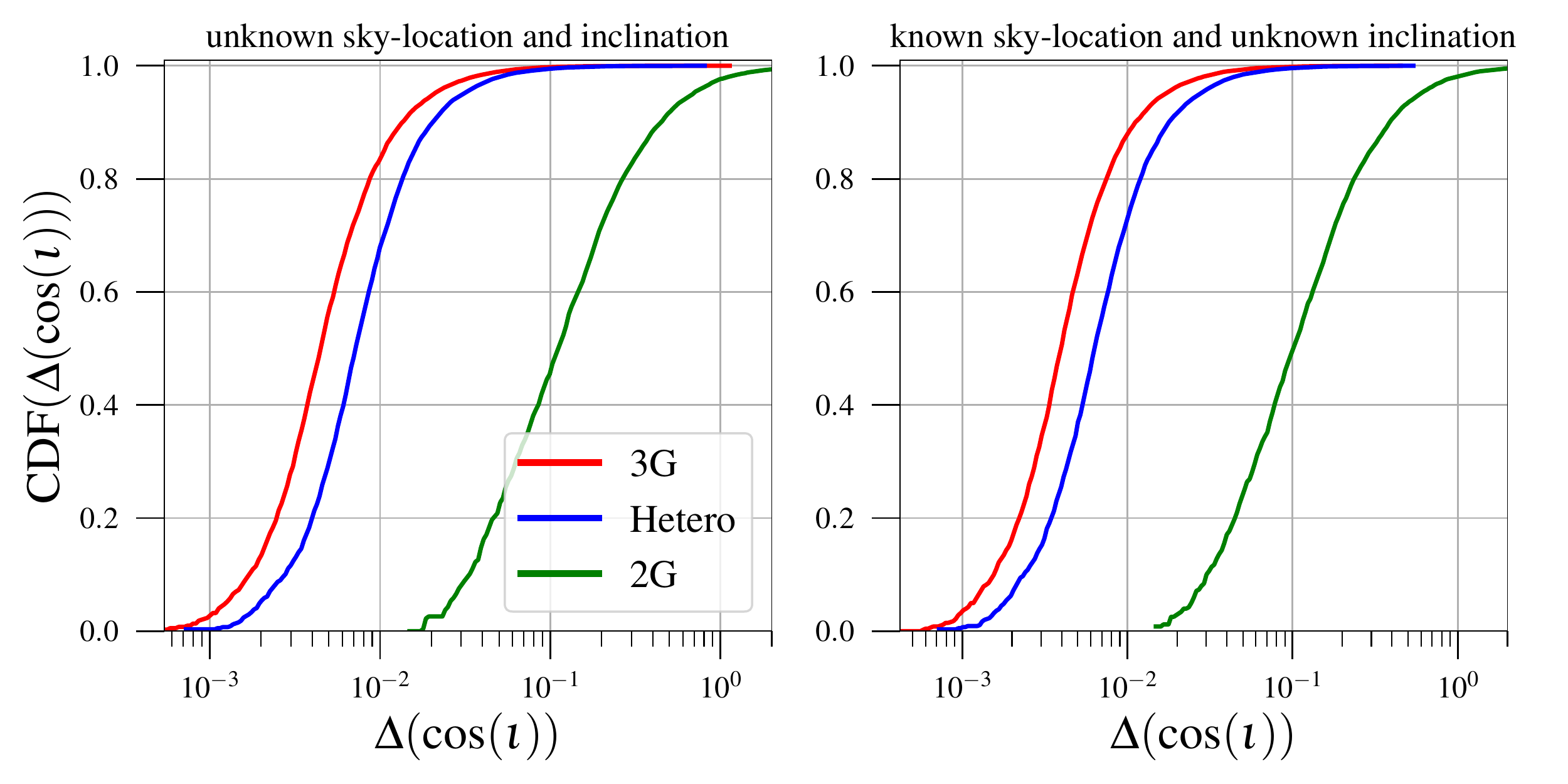}
\caption{Cumulative distribution of $1-\sigma$ inclination errors measured with various networks of detectors, 2G, 3G, Hetero, summarized in Tab.~\ref{tab_det_net_info}. 
A population of binary neutron stars with fixed masses $m_1 = 1.45M_{\odot}$ and $m_2 = 1.35M_{\odot}$ have isotropic sky-locations and orbital inclinations and 
are uniformly distributed in the co-moving volume between 10 Mpc and 300 Mpc.
Left panel shows the errors when sky-location and orbital inclination of the binaries are not known to us. Right panel shows the 
error when sky-location of the binaries are known. All the sources plotted here have network SNR $\geq10$.}
\label{fig:iota}
\end{figure*}
\begin{figure}[h!]\centering
\includegraphics[scale=0.55]{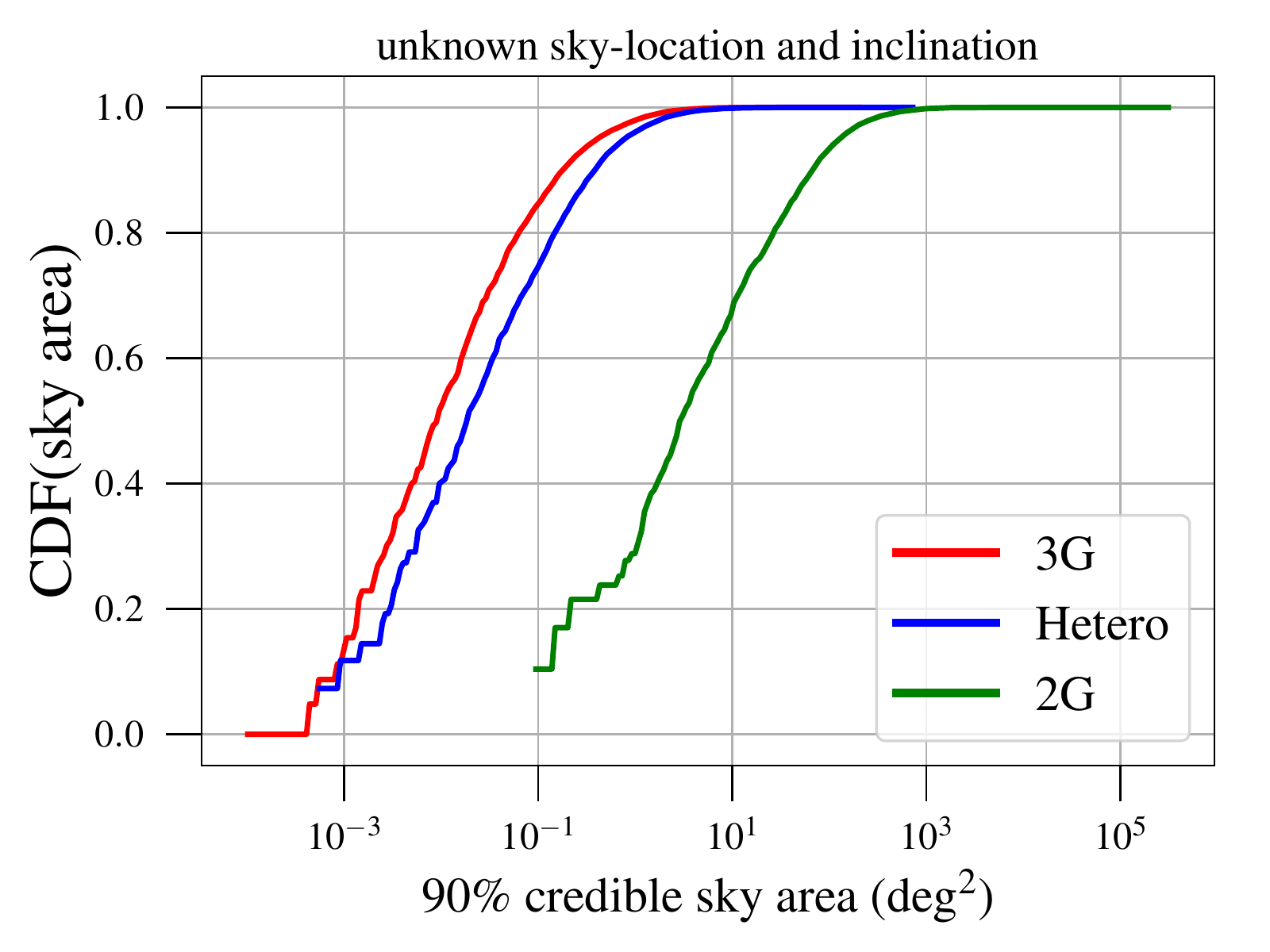}
\caption{Cumulative distribution of $90\%$ credible sky-area measured with various networks of detectors, 2G, 3G, Hetero, summarized in Tab.~\ref{tab_det_net_info}. 
A population of binary neutron stars with fixed masses $m_1 = 1.45M_{\odot}$ and $m_2 = 1.35M_{\odot}$ have isotropic sky-locations and orbital inclinations and 
are uniformly distributed in the co-moving volume between 10 Mpc and 300 Mpc. All the sources plotted here have network SNR $\geq10$.}
\label{fig:skyarea}
\end{figure}

Currently we have three second generation (2G) GW detectors that are operational: advanced LIGO (aLIGO) in Hanford-USA, aLIGO in Livingston-USA, and advanced Virgo (AdV) in Italy \citep{aLIGO_ref, AdV_ref}. The Japanese detector KAGRA \citep{KAGRA_ref, Somiya:2011np} is expected to join the network in the third observing 
run, and the detector in the Indian continent, LIGO-India, is expected to be online by 2025 \citep{Ligo-india}. Therefore, in a few years time we will have a network of 2G detectors fully operational, observing the GW sky. We call such a network of second generation detectors the ``2G network''. At present, significant efforts are on-going to put forward the science case for the third generation (3G) GW detectors such as Cosmic explorer (CE) \citep{CE_ref} and Einstein telescope (ET) \citep{ET_ref}. These 3G detectors will not only let us `hear' deeper in the universe, allowing more and more detections, but will also help us study each source in great detail. These 3G detectors are expected to be online sometime in 2030s. Therefore, by that time we will have a network of 3G detectors, say, ET in Italy, one CE in Utah-USA and another CE in Australia. It has been found that by placing 3G detectors on the globe in this manner, we will be able to achieve maximum science goals \citep{Hall2019}. We term such a network of detectors as ``3G network''. Furthermore, there are also plans to improve the sensitivity of existing detectors at LIGO sites by a factor two by using high power lasers and better and bigger test masses, these are called `LIGO Voyager'\footnote{https://dcc.ligo.org/LIGO-T1500290/public}. Given this we will have LIGO Voyager, as well, by the time 3G detectors come online. Therefore, we assume a hypothetical network of detectors constituting 3G and Voyager detectors: CE in Utah-USA, one Voyager in Livingston-USA, ET in Italy, one Voyager in India and one Voyager in Japan, and we name this as ``Heterogeneous network''. Table \ref{tab_det_net_info} lists the detector networks used in this paper to measure binary distances, along with their location on Earth and the associated noise sensitivity curves\footnote{We use the an analytical fit given in \citet{Ajith2011b} for the power spectral density (PSD) of aLIGO. The PSD for AdV, KAGRA and Voyager are taken from \texttt{https://dcc.ligo.org/LIGO-T1500293/public}. For ET we use the data given in \citet{ET-D} and for CE we use the analytical fit given in \citet{Kastha2018}.}. 
Figure~\ref{fig:snr} presents the cumulative distribution of network signal-to-noise ratios (SNRs) for the binary neutron star population we considered in this paper while using 2G, 3G and Hetero networks.

To measure the errors in the distance we use the {\it Fisher information matrix} technique \citep{Rao45,Cramer46}. This is a useful semi-analytic method that employs a
quadratic fit to the log-likelihood function and derives $1-\sigma$ error bars on 
the binary parameters from its GW signal~\citep{CF94,AISS05}. 
Given a frequency-domain GW signal $\tilde{h}(f; \vek \theta)$, described by the
set of parameters $\vek \theta$, the Fisher information matrix is given as
\begin{equation}
\Gamma_{ij}=\langle\tilde{h}_i,\tilde{h}_j\rangle,
\label{eq:fisher}
\end{equation}
where $\tilde{h}_i=\partial\tilde{h}(f;\vek \theta)/\partial\theta_i$, and the angular bracket, $\langle...,...\rangle$,  denotes the noise-weighted inner product 
defined by 
\begin{equation}
\langle a,b\rangle=2\int_{f_{\rm low}}^{\rm f_{\rm high}}\frac{a(f)\,b^*(f)+a^*(f)\, b(f)}{S_h(f)}\,df \,.
\label{eq:innerproduct}
\end{equation}
Here $S_h(f)$ is the one-sided noise power spectral density (PSD) of the detector and $[f_{\rm low}, f_{\rm high}]$ are the limits of integration.
The variance-covariance matrix is defined by the inverse of the Fisher 
matrix, $C^{ij}=(\Gamma^{-1})^{ij},$ where the diagonal components, $C^{ii}$, are 
the variances of $\theta_i$. The $1-\sigma$ errors on $\theta_i$ is, therefore, given as
\begin{equation}
\Delta \theta_i = \sqrt{C^{ii}} \,.
\end{equation}  
In the case of a network of detectors, one computes Fisher matrices $\Gamma^A$ corresponding to each detector $A$ and adds them up
\begin{equation}
\Gamma^{\rm net} = \sum_A \Gamma^A \,.
\end{equation}
The error in the parameters is then given as  $\Delta \theta_i = \sqrt{C^{ii}}$ where $C$ is now the inverse of $\Gamma^{\rm net}$.

As the chirp mass, ${\cal M} = (m_1m_2)^{3/5}/(m_1+m_2)^{1/5}$ and symmetric mass ratio, $\eta = m_1m_2/(m_1+m_2)^2$ 
are the best measured mass parameters by GW observations during the inspiral phase of a binary, 
we assume our parameter space to be $\vek \theta = \{{\rm ln}{\cal M}, {\rm ln}\eta, {\rm ln} D_L, \cos(\iota), \cos(\theta), \phi, \psi, t_c, \phi_c \}$. 
Fisher matrix based parameter estimation in the context of third generation detectors have been done in the past \citep{Wen2018,Leong2018}.
In this paper, we compute fractional error in the distance measurement, $\Delta D_L/D_L$, using the 
detector networks listed in Tab.~\ref{tab_det_net_info}, and 
the results in various observational scenarios are as follows: 

\paragraph{ Unknown sky position and inclination}: In this scenario, we assume that nothing is known about the binaries and compute errors in all the parameters using 9-dimensional Fisher matrix. This scenario is relevant when we can not identify the electromagnetic counterpart of the binary neutron stars and all the information about the source is coming from GW observation alone. We compute $1-\sigma$ error in the parameters $\{{\rm ln}{\cal M}, {\rm ln}\eta, {\rm ln} D_L, \cos(\iota), \cos(\theta), \phi, \psi, t_c, \phi_c \}$ and the cumulative distribution of fractional error in the distance measurement,  $\Delta D_L/D_L$, is shown on the right most panel of Fig.~\ref{fig:dist}. 
 We observe that the 3G network performs slightly better than the Hetero network, constraining distances with a median of $\sim 1.6\%$ accuracy ($90\%$ sources have error $\lesssim3\%$). The reason behind 3G network performing better than Hetero is because  3G network has 3 third generation detectors whereas Hetero contains only 2 such detectors. The network of second generation detectors, on the other hand, performs very poorly providing distance estimates with $\sim 50\%$ error ($90\%$ sources have error $\lesssim60\%$). On the left panel of Fig.~\ref{fig:iota}, we present the distribution of $1-\sigma$ error in the measurement of cosine of the inclination angle $\iota$. Again, 3G and Hetero networks achieve similar accuracies with a median error of  $\sim 0.01$ whereas 2G network  performs an order of magnitude worse, constraining $\cos \iota$ with median error of  $0.4$.  Figure~\ref{fig:skyarea} presents the cumulative distribution of $90\%$ credible area of binaries on the sky. The 3G network gives the best estimate for the sky location followed by Hetero network. For instance, the 3G network will be able to locate binary neutron star merger (on an average) within $\sim1$ deg$^2$ whereas the Hetero network can have the $90\%$ credible sky area $\sim1.4$ deg$^2$, and the 2G network could only pinpoint the binary neutron stars with $\sim 180$ deg$^2$ sky-area. 

\paragraph{Known sky position but unknown inclination}: In this scenario, we assume that the sky position of the binary neutron stars are known through their electromagnetic observations. We, therefore, use the information of $\theta$ and $\phi$ of the sources and compute only 7-dimensional Fisher matrix for parameters:  $\{{\rm ln}{\cal M}, {\rm ln}\eta, {\rm ln} D_L, \cos(\iota), \psi, t_c, \phi_c \}$. The cumulative distribution of error in the distance measurement is shown in the middle panel of Fig.~\ref{fig:dist} and we notice that the accuracy has slightly improved now for all the networks. This is because the knowledge of source's sky position breaks down the degeneracy between the sky-location angles ($\theta, \phi$) and distance $D_L$ and allow us to measure source distance relatively better. The 3G and Hetero networks are still performing far better than the 2G network. The right panel of Fig.~\ref{fig:skyarea} shows the distribution of error in $\cos (\iota)$ and it has slightly improved as compared to the case when the sky-position of the source is not known. 

\paragraph{Known sky-position and inclination}: This scenario assumes that the sky-position as well as the inclination angles of the binary neutron stars are known purely from their electromagnetic counterparts. This scenario is possible as we already have seen in the case of GW170817. The sky position of GW170817 was constrained by finding the host galaxy NGC 4993 through numerous optical and infrared observations \citep{GBM:2017lvd} whereas the inclination angle or the so-called ``opening angle'' was constrained from the X-ray and ultraviolet observations \citep{Evans:2017mmy}. This scenario has a merit as the error in the distance measurement can be significantly reduced as shown in the right most panel of Fig.~\ref{fig:dist}. In this scenario, we use the information of $\theta$, $\phi$ and $\iota$ and compute 6-dimensional Fisher matrices for parameters, $\{{\rm ln}{\cal M}, {\rm ln}\eta, {\rm ln} D_L, \psi, t_c, \phi_c \}$.
All the degeneracies between the distance $D_L$ and $\theta$, $\phi$ and $\iota$ are now broken which give us highly accurate distance measurement with median error of $\sim0.5\%$ for 3G and Hetero networks ($90\%$ sources have error $<0.8\%$).  

\begin{figure*}
\centering
\includegraphics[scale=0.8]{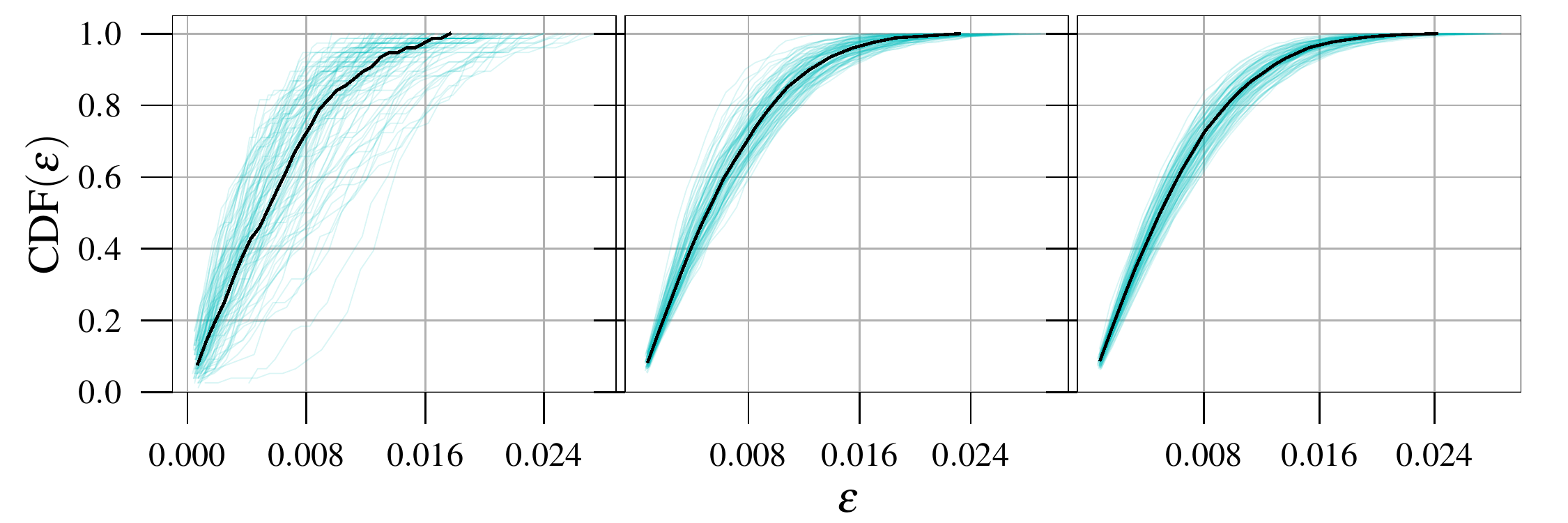}
\caption{Cumulative distribution of $\epsilon$, the fractional difference
 between binary neutron star mergers and SNe Ia distances in Coma cluster. The cyan curves are
100 realization of sampling radial positions of galaxies in Coma using \texttt{halotools} and the black curve 
represents the median. Left, middle and right panels assume that there are 2, 13 and 27 binary neutron star mergers in 
Coma, respectively.}
\label{fig:epsilon_r}
\end{figure*}

Given the measurement capabilities of the different detector networks we can now assess whether it will be possible to localize a merger event uniquely to a galaxy cluster. As we shall argue unique identification of a galaxy cluster associated with a binary neutron star merger will be possible in a 3G or a heterogeneous network for 80\% of the sources. From Fig.\,\ref{fig:dist}, left panel, we see that in the 3G (heterogeneous) network, for 80\% of binary mergers the 90\% credible interval in the measurement of the luminosity distance is 2\% (respectively, 3\%) at distances up to 300 Mpc. The corresponding 90\% uncertainty in the sky position of the source is $\sim 0.1$ square degrees for both 3G and heterogeneous network (see Fig.\,\ref{fig:skyarea}), with the 3G network performing slightly better. These numbers correspond to a maximum error in distance of $\Delta D_L \sim 9\,\rm Mpc$ and an angular uncertainty of $\Delta\Omega \sim 3\times 10^{-5}\,\rm str,$ which correspond to an error box in the sky of 
$$\Delta V \simeq D_L^2 \Delta D_L \Delta \Omega \simeq 25\,{\rm Mpc^3} \left ( \frac{D_L}{300\,\rm Mpc} \right )^2.$$
Given that the number density of galaxies is $3\times 10^6\,\rm Gpc^{-3}$, the error box $\Delta V$ will contain no more than one field galaxy; if the merger occurs in a cluster, it will be localized to a unique cluster as the number density of clusters is far smaller than those of field galaxies. However, without an electromagnetic counterpart it will not be possible to associate a merger to a unique galaxy within a cluster, as the number density of galaxies in a cluster will be far greater than the number density of field galaxies. 

In summary, given that we have restricted our analysis to rich clusters that are a sixth of Coma or larger, gravitational wave observations alone will associate most mergers in clusters to a unique galaxy cluster; an electromagnetic counterpart will be needed to further associate the event to a specific galaxy within a cluster.

\section{Calibrating TYPE Ia Supernovae with Binary Neutron Star Mergers in a Galaxy Cluster}
\label{sec:calibrate}

When a binary neutron star merger event occurs in a galaxy cluster we may have tens of SNe Ia in the same cluster. How do we calibrate SNe Ia in one of these galaxies given the distance to the host galaxy of the binary merger? The problem is that we would not know the relative positions of SNe Ia and binary merger host galaxy.  In this section we derive the distribution of the error one would make if one assumed that both transients occurred in the same galaxy. In other words, we investigate how the dispersion of galaxies throughout the cluster might affect the distance estimation of 
SNe Ia calibrated through GW events in the same cluster. An additional source of error arises from the peculiar velocity of host galaxies of the transient events. In the second part of this section we provide a rough estimate of how large this effect might be. 

\paragraph{Error due to position uncertainty of SNe Ia hosts:}
To this end, we take the example of the Coma cluster. The Coma cluster 
is roughly 100 Mpc away from Earth and contains more than 3000 galaxies.  
Following several studies \citep{Lokas:2003ks, Brilenkov:2015uxa} we assume that the matter density 
in Coma can be well 
approximated by the Navarro-Frenk-White profile \citep{Navarro:1995iw}. To simulate positions of galaxies within this
cluster we use the publicly available python-package \texttt{halotools} \citep{Hearin:2016uxs} which requires the number of galaxies in a cluster, their {\em concentration,} and the mass of the cluster as input parameters. We simulate 1000 galaxies and assume the concentration and mass of the cluster to be  4 and $1.29\times10^{15}M_{\odot}\, h^{-1}$, respectively, as reported in \citet{Brilenkov:2015uxa}. We consider $h$ to be $0.701$. 

In Sec.~\ref{sec:rates}, we learned that ten years of optical observation would allow us to calibrate
roughly 38 SNe Ia per binary neutron star merger host galaxy cluster. Furthermore, we expect
to observe between 1.8 and 26.6 such clusters within 300 Mpc in five years of GW observation period. 
For simplicity in our calculations, we assume that all these clusters are Coma-like, i.e., they all have same 
matter density profile and each contains 1000 galaxies. Let us consider that one detects a binary neutron star merger 
in a particular galaxy cluster, it will then be 
accompanied by $38$ SNe Ia within 10 years of optical observation. 
We distribute 1 binary neutron star and 38 SNe Ia 
randomly among cluster's 1000 simulated galaxies, and calculate the fractional difference $\epsilon$ 
in the luminosity distances of binary neutron star merger and SNe Ia as
\begin{equation}
\epsilon = \frac{|D_{\rm BNS}-D_{\rm SNeIa}|}{D_{\rm BNS}} \,,
\end{equation}
where $D_{\rm BNS}$ and $D_{\rm SNeIa}$ are the true distances of binary neutron star mergers and SNe Ia, respectively, in our simulation. With one galaxy cluster we obtain 38 samples of $\epsilon$, and since all the 
 clusters are the same it is easy to scale this number with the number of clusters. More explicitly, 
having two clusters with each containing 1 binary neutron star merger and 38 SNe Ia is equivalent to 
have one cluster containing 2 binary neutron star mergers and 76 SNe Ia. 
Following this argument, in Fig.~\ref{fig:epsilon_r} we plot the cumulative distribution 
of $\epsilon$ for 2, 13 and 27 binary neutron star mergers in a cluster (we round the number of clusters to the nearest integer).
The cyan colors show 100 realization of sampling radial positions of galaxies 
in Coma using \texttt{halotools} and the black curve represents the median.   
From Fig.~\ref{fig:epsilon_r} we note that 90\% (99\%) of the times $\epsilon < 0.9\%$ ($<1.5\%$) 
which implies that there will be ${\cal O}(1\%)$ error in the distance estimation of SNe Ia if calibrated through binary 
neutron star mergers in the same galaxy cluster. 
\begin{figure}[hb]
\centering
\includegraphics[width=0.45\textwidth]{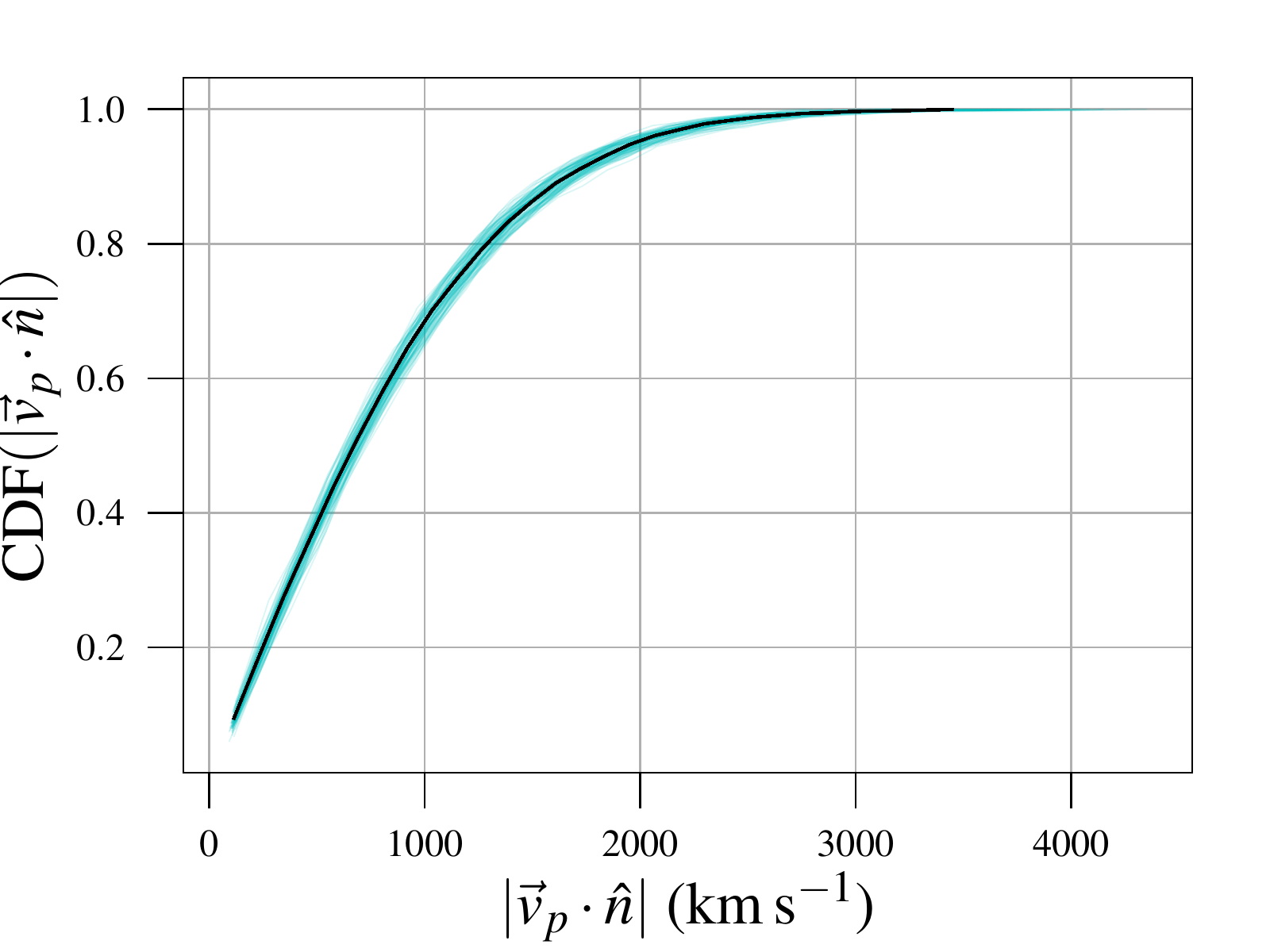}
\caption{Cumulative distribution of magnitude of the 
line of sight peculiar velocity, $|\vec v_p \cdot \hat n|$, of galaxies in the Coma cluster. The cyan curves are
100 realization of sampling radial velocities of galaxies in Coma using \texttt{halotools} and the black curve 
represents the median.} 
\label{fig:epsilon_v}
\end{figure}

\paragraph{Error due to peculiar velocities of host galaxies:}
In a rich cluster, galaxies can have quite a large peculiar velocity. For example, \citet{Lokas:2003ks} quote that the peculiar velocity $\vec v_p$ in the Coma cluster can be as large as $\sim10^4\,\rm km\,s^{-1}$, while typical rich clusters are known to have $|v_p|\sim 750\,\rm km\,s^{-1}$ \citep{Bahcall:1995tf}.  What is relevant is the peculiar velocity projected along the line-of-sight $\hat n,$ namely $\vec v_p \cdot \hat n$, because it is this velocity that affects the apparent luminosity of SNe Ia and binary neutron star mergers due to the Doppler effect.  For $\vec v_p$ of a constant magnitude but distributed isotropically in space we would expect the line of sight RMS velocity to be $\vec v_p/\sqrt{3}.$ However, $\vec v_p$ varies throughout the cluster, and for Coma using \texttt{halotools} we find $\overline v \equiv \langle (\vec v_p \cdot \hat n)^2 \rangle^{1/2} \sim 10^{3}\,\rm km\,s^{-1}$, as shown in Fig~\ref{fig:epsilon_v}, where $\langle \dots \rangle$ stands for average over all directions. 

The luminosity distance inferred to a binary system is affected by the local peculiar velocity.
The error induced in the luminosity distance due to the RMS line-of-sight velocity $\overline v$ is 
${\delta D_L} = \overline v/H_0.$ Hence, for $H_0 = 70\,\rm km\,s^{-1}\,Mpc^{-1},$ the error in binary's distance is
${\delta D_L} \simeq 14\,\rm Mpc$. This is the typical error we make in the estimation of distance due to peculiar velocity and it remains the same for a cluster of given concentration. Thus, at the distance of the Coma cluster, this error is $\sim 14\%$ while it reduces to $\sim 5\%$ for clusters at 300 Mpc. As seen in Fig.~\ref{fig:dist}, the error in luminosity distance of binaries due to GW measurements alone (assuming that the host's sky position is known) is $\sim 1.2 \%,$ which is far less compared to the error due to peculiar motion. However, it is comparable to the error due to the position uncertainty relative to binary neutron star merger of SNe Ia that we discussed above. Thus, the calibration uncertainty of SNe Ia up to 300 Mpc is largely due to the peculiar motion of galaxies.

\begin{figure}[h!]\centering
\includegraphics[scale=0.55]{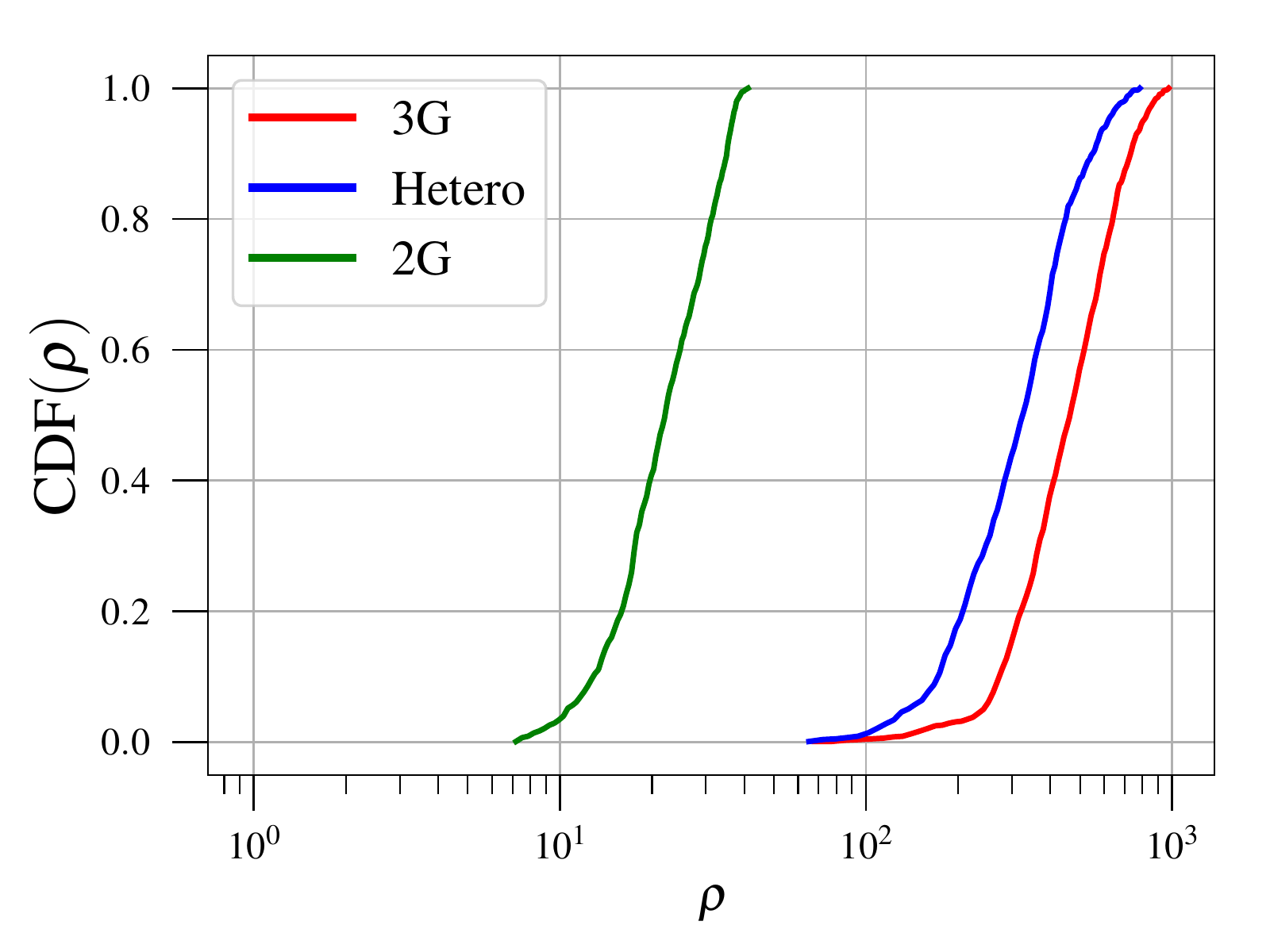}
\caption{Cumulative distribution of network SNR of BNSs in galaxies in Coma cluster measured with various networks of detectors. 
The binary neutron stars in these galaxies have fixed masses $m_1 = 1.45M_{\odot}$ and $m_2 = 1.35M_{\odot}$ and isotropic orbital inclinations.}
\label{fig:snr_coma}
\end{figure}

\begin{figure*}[h!]\centering
\includegraphics[scale=0.48]{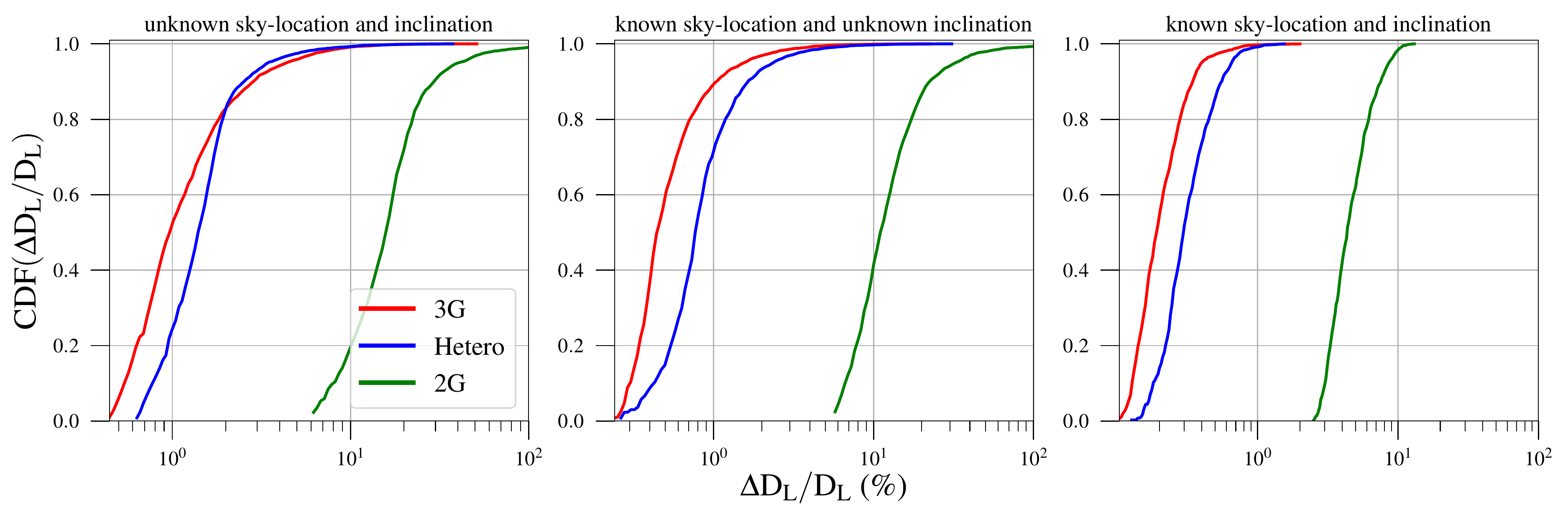}
\caption{Cumulative distribution of $1-\sigma$ distance errors of galaxies in Coma cluster measured with various networks of detectors. 
The binary neutron stars in these galaxies have fixed masses $m_1 = 1.45M_{\odot}$ and $m_2 = 1.35M_{\odot}$ and isotropic orbital inclinations.
Left panel shows the errors when sky-location and orbital inclination of the binaries are not known to us. Middle panel shows the 
error when sky-location of the binaries are known and the right panel demonstrates distance errors when both sky-location and orbital inclination of binaries are 
known. All the sources plotted here have network SNR $\geq10$.}
\label{fig:dist_coma}
\end{figure*}

\begin{figure*}[h!]\centering
\includegraphics[scale=0.65]{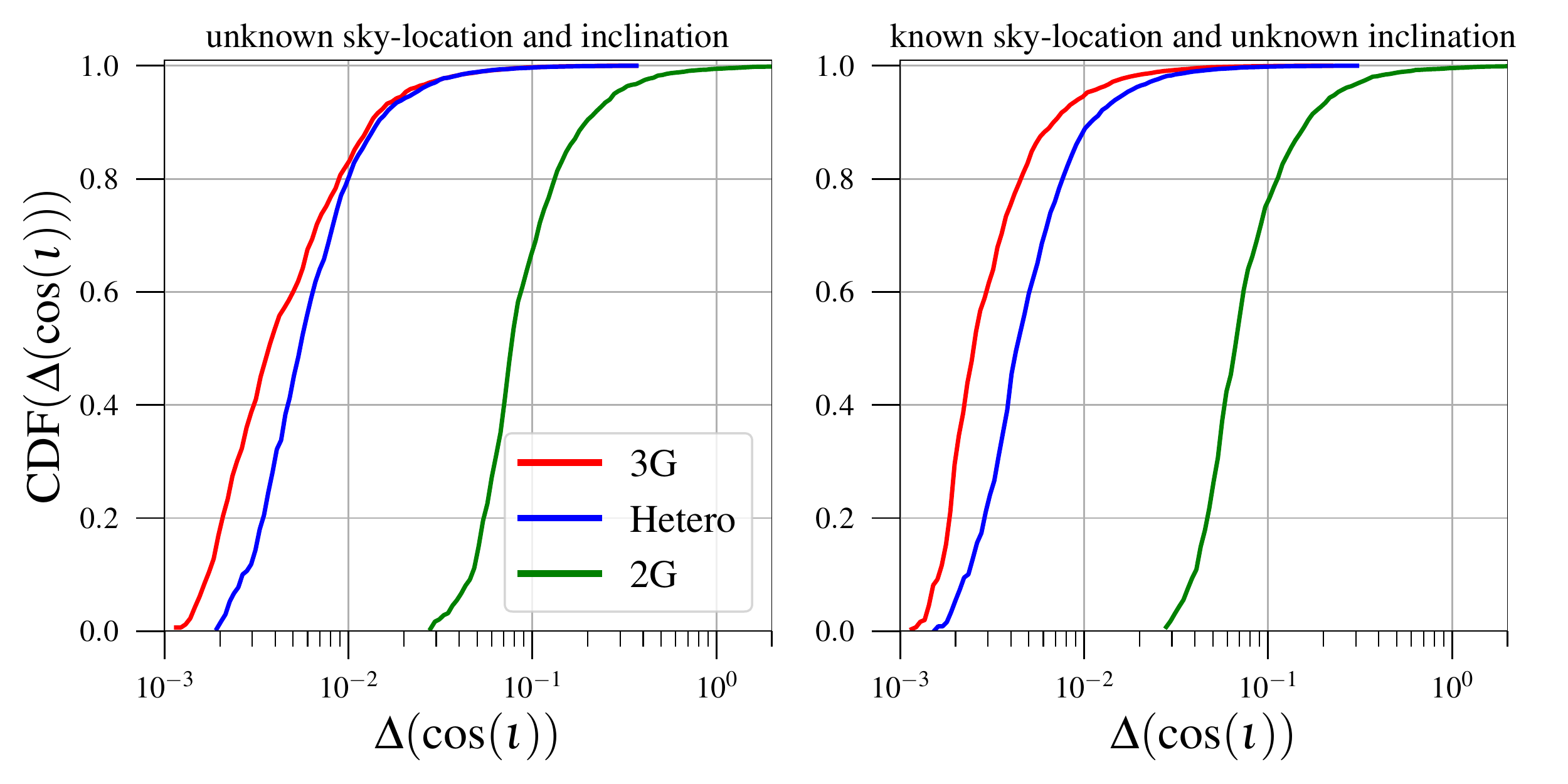}
\caption{Cumulative distribution of $1-\sigma$ errors in measurement of orbital inclination of binary neutron stars residing in galaxies in Coma cluster. 
The binary neutron stars in these galaxies have fixed masses $m_1 = 1.45M_{\odot}$ and $m_2 = 1.35M_{\odot}$ and isotropic orbital inclinations.
Left panel shows the errors when sky-location and orbital inclination of the binaries are not known to us. Right panel shows the 
error when sky-location of the binaries are known. All the sources plotted here have network SNR $\geq10$.}
\label{fig:iota_coma}
\end{figure*}

\begin{figure}[h!]\centering
\includegraphics[scale=0.55]{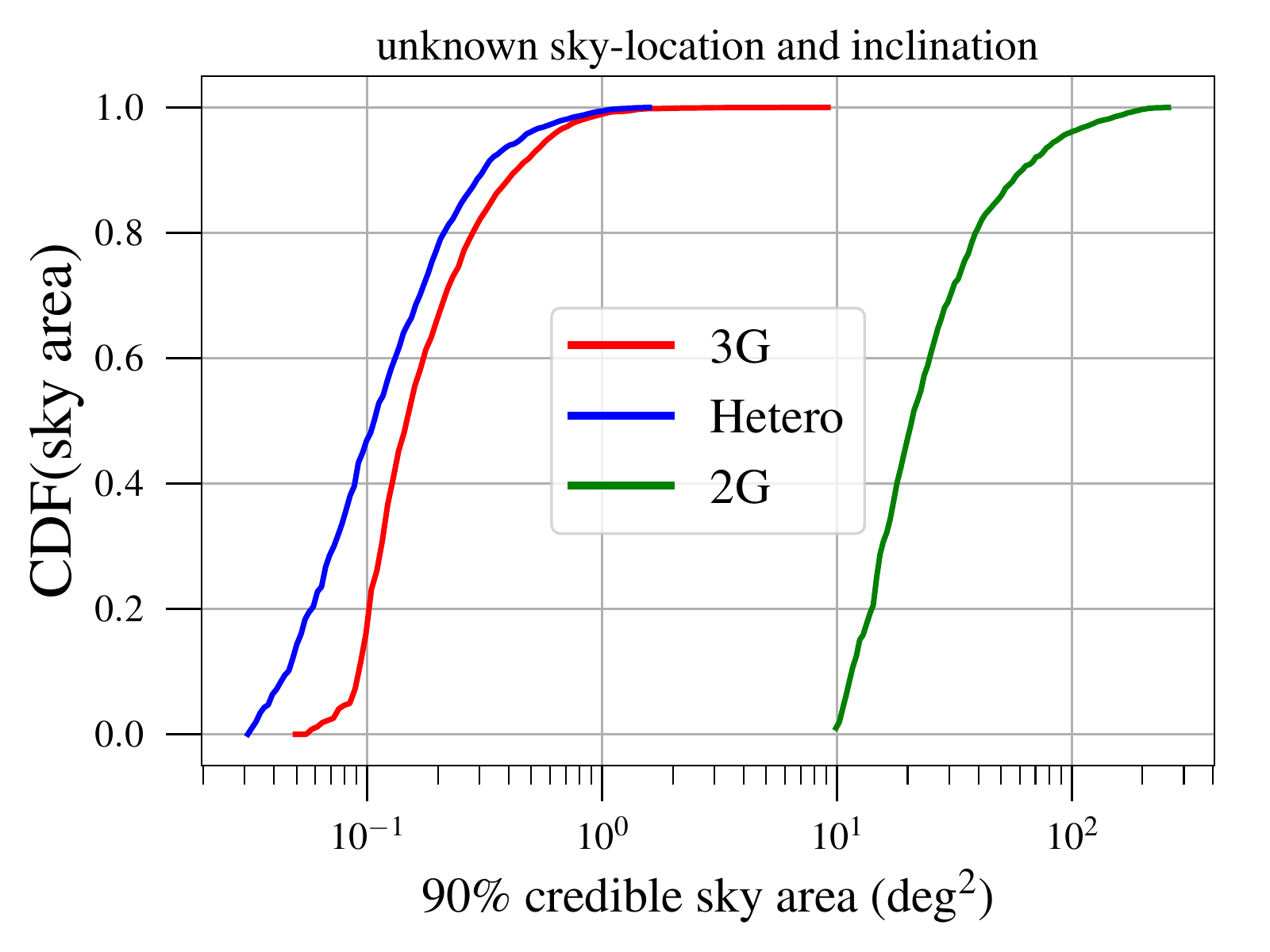}
\caption{Cumulative distribution of $90\%$ credible sky-area of galaxies in Coma cluster measured with various networks of detectors. 
The binary neutron stars in these galaxies have fixed masses $m_1 = 1.45M_{\odot}$ and $m_2 = 1.35M_{\odot}$ and isotropic sky-locations and orbital inclinations. All the sources plotted here have network SNR $\geq10$.}
\label{fig:skyarea_coma}
\end{figure}

However, what is the typical error in the distance 
measurement of the binary merger itself in these Coma-like clusters? 
We compute the error in the distance measurement of galaxies in Coma using different networks of 
detectors.\footnote{In order to sample the sky positions with respect to Earth, we assume that
the center of Coma cluster is located on the sky with $\theta=27.98^{\circ}$ and $\phi= 194.95^{\circ}$.}
Figure~\ref{fig:snr_coma} shows the cumulative distribution of network SNR for this population of binary neutron stars in Coma for 2G, 3G and Hetero detector networks.
We compute the error in binary's distance measurement in all the three observational scenarios we discussed in the previous section and the results are shown in Fig.~\ref{fig:dist_coma}. The 3G network performs the best in constraining distances with median of $\sim 2\%$ error ($90\%$ sources have error $<3\%$) when the electromagnetic counterpart of the binary neutron star merger can not be identified. The error reduces to $\sim 0.3\%$ ($90\%$ sources have error $<0.4\%$) when both the sky-position and inclination angle are known from the electromagnetic observations. Figure~\ref{fig:iota_coma} and  \ref{fig:skyarea_coma} depict the cumulative distribution of errors in the measurement of $\cos(\iota)$ and $90\%$ credible sky area, respectively. 

This shows that the error in the estimation of SNe Ia distance due to GW calibration is comparable to the statistical error in the measurement of the calibrator's distance itself for the galaxies in the Coma cluster.  
 
\section{Discussion: Gravitational Waves as A Cosmic Distance Ladder}
\label{sec:new_rung}
In this paper we explored the possibility of calibrating type Ia
supernovae using gravitational waves from coalescing binary neutron
stars as standard sirens. According to the current best estimates, the
volumetric rate of SNe Ia is 30 times larger than binary neutron star
mergers. Even so, there is a very little chance that a SNe Ia would
occur in the same galaxy as a binary neutron star merger. However,
when a neutron star merger occurs in a galaxy cluster it is guaranteed
that more than one SNe Ia would have occurred in the same
cluster within a year. As shown in Fig. \ref{fig:clsim} in a typical rich cluster within 300 Mpc, 
such as Coma, a binary neutron star merger will be accompanied by a few SNe Ia
each year, providing ample opportunity to calibrate supernovae using
standard sirens.

To accomplish this task it is necessary to control the error in the
measurement of distance to merging binary neutron stars to well below the
other sources of error, such as the unknown relative positions of SNe
Ia and the peculiar velocity of galaxies within a cluster. 
One makes an error of $\sim$0.9\% in distance of SNe Ia, for 90\% of the supernovae,
when one does not know the host galaxies of either SNe Ia or binary merger in
a Coma-like cluster and assumes both of them to occur in the same galaxy.  
On the other hand, one makes an error of $\sim$14\% due to the peculiar velocities of 
galaxies in the Coma-like cluster. Note that Coma is 100 Mpc away from Earth and both these
errors translate to $\sim$0.3\% and $\sim$5\%, respectively, for
galaxies at 300~Mpc. 
In contrast, we find that the next generation of
GW detector network (one Einstein Telescope and two Cosmic Explorers)
will be able to obtain distance error for the standard sirens to be
less than 1\% for 90\% of the binary neutron star mergers whose sky
position and inclination are known from electromagnetic observations
within 300~Mpc.  Thus, the prospect of calibrating SNe~Ia using a
completely independent method and establishing a new cosmic distance
ladder looks bright. 

SNe~Ia are expected to remain a key tool for distance estimation and
cosmology through the next decade and beyond. A particularly exciting
near-term prospect is the ten-year LSST survey \citep{lsst_2017}, due
to begin by the end of 2022. LSST will discover and characterize $\sim$50,000
SNe~Ia per year out to redshift $z\approx 0.7$ in its main survey
fields, and an additional $\sim$1500 per year out to redshift
$z\approx 1.1$ in its ``deep drilling'' fields; of these SNe Ia,
$\approx$200 per year are anticipated to have LSST data of sufficient
quality to support cosmological analyses.
Although spectroscopic characterization of all but a fraction of LSST SNe~Ia
will not be feasible, photometric analyses of the SNe Ia and their host
galaxies, in the context of the sheer number of events, are expected
to enable high-quality constraints on cosmology, particularly the
matter density $\Omega_m$ and Dark Energy equation of state $w$. (For
LSST's ultimate cosmological studies, the SNe~Ia analysis will be
combined with weak lensing measurements of mass clustering and the
growth of structure, and a cosmic scale factor analysis from the
baryon acoustic oscillations feature of large scale structure, to
yield joint constraints on all cosmological parameters.)

A GW-based calibration of the LSST sample of SNe~Ia can be achieved at
low redshift via binary neutron star detections from the jointly-observed redshift
range $0.02\le z \le 0.07$ ($85\,{\rm Mpc}\lesssim D_L \lesssim 300\,{\rm
Mpc}$). Over this range, binary neutron star mergers will be detectable by
next-generation GW facilities, while at the same time the effects of
galaxy peculiar velocities will be minimal ($<$5\% per object for
field galaxies). LSST simulations \citep{lsst_2017} project
high-quality characterization of $\approx200$ SNe~Ia per year in this
redshift range, and the estimated binary neutron star merger rates are 12 to 420
(median 110) per year for this 0.11~Gpc$^{3}$ volume. This
suggests that a high-quality GW-based calibration of SNe~Ia
luminosities in the field should also be possible in the LSST era.

In conclusion, the fundamental advance considered in this paper is
provided by the application of precision GW-based distance
measurements \citep{Schutz86} to the calibration of type~Ia SN
distances -- specifically, in cases where events of both types are
hosted by a single galaxy cluster. Considering the broader picture,
the impending realization of a longstanding astronomical dream of
precise distance estimates on near-cosmological scales can be expected
to yield many additional applications. For example: Precision studies
of galaxy and galaxy cluster peculiar velocities; three-dimensional
mapping of galaxies in the context of their host clusters and groups;
and the fully tomographic use of galaxies and active galactic nuclei to characterize the
gas, stellar, and dark matter contents of their host groups and
clusters. Given the implications of precise distance measurements for
nearly every branch of astronomy and astrophysics, a mere refinement
of our present understandings would be in some sense a disappointment.
We choose to hope, instead, for at least a few genuine surprises.

\section*{Acknowledgments}
We thank  Andrew Hearin, Aseem Paranjape,  Robin Ciardullo and Rahul Srinivasan
for useful discussions. We thank Christopher Messenger for carefully reading the manuscript 
and providing useful comments. We also thank the anonymous referee for their critical comments which significantly improved the presentation of this manuscript. 
AG and BSS are supported in part by NSF grants
PHY-1836779, AST-1716394 and AST-1708146. BSS and BFS gratefully 
acknowledge support from the Science and Technology Facilities Council (STFC) of the United Kingdom.
We acknowledge the use of IUCAA LDG cluster Sarathi for the 
computational/numerical work. This paper has the LIGO document number LIGO-P1900172.


\begin{thebibliography}{}
\expandafter\ifx\csname natexlab\endcsname\relax\def\natexlab#1{#1}\fi

\bibitem[{Aasi {et~al.}(2015)}]{aLIGO_ref}
Aasi, J., {et~al.} 2015, Class. Quant. Grav., 32, 074001

\bibitem[{Abbott {et~al.}(2017{\natexlab{a}})Abbott, Abbott, Abbott, Abernathy,
  Ackley, Adams, Addesso, Adhikari, Adya, Affeldt, Aggarwal, Aguiar, Ain,
  Ajith, Allen, Altin, Anderson, Anderson, Arai, Araya, Arceneaux, Areeda,
  Arun, Ashton, Ast, Aston, Aufmuth, Aulbert, Babak, Baker, Ballmer, Barayoga,
  Barclay, Barish, Barker, Barr, Barsotti, Bartlett, Bartos, Bassiri, Batch,
  Baune, Bell, Berger, Bergmann, Berry, Betzwieser, Bhagwat, Bhandare, Mandic,
  \& {ligo Scientific Collaboration}}]{ET-D}
Abbott, B., Abbott, R., Abbott, T., {et~al.} 2017{\natexlab{a}}, Classical and
  Quantum Gravity, 34, doi:10.1088/1361-6382/aa51f4

\bibitem[{Abbott {et~al.}(2017{\natexlab{b}})}]{GW170817}
Abbott, B., {et~al.} 2017{\natexlab{b}}, Phys. Rev. Lett., 119, 161101

\bibitem[{Abbott {et~al.}(2016{\natexlab{a}})}]{GW151226}
Abbott, B.~P., {et~al.} 2016{\natexlab{a}}, Phys. Rev. Lett., 116, 241103

\bibitem[{Abbott {et~al.}(2016{\natexlab{b}})}]{GW150914}
---. 2016{\natexlab{b}}, Phys. Rev. Lett., 116, 061102

\bibitem[{Abbott {et~al.}(2017{\natexlab{c}})}]{GW170817_H0}
---. 2017{\natexlab{c}}, Nature, 551, 85

\bibitem[{Abbott {et~al.}(2017{\natexlab{d}})}]{Abbott:2016jsd}
---. 2017{\natexlab{d}}, Phys. Rev., D95, 062003

\bibitem[{Abbott {et~al.}(2017{\natexlab{e}})}]{CE_ref}
---. 2017{\natexlab{e}}, Class. Quant. Grav., 34, 044001

\bibitem[{Abbott {et~al.}(2017{\natexlab{f}})}]{GW170104}
---. 2017{\natexlab{f}}, Phys. Rev. Lett., 118, 221101

\bibitem[{Abbott {et~al.}(2017{\natexlab{g}})}]{GW170608}
---. 2017{\natexlab{g}}, Astrophys. J., 851, L35

\bibitem[{Abbott {et~al.}(2017{\natexlab{h}})}]{GW170814}
---. 2017{\natexlab{h}}, Phys. Rev. Lett., 119, 141101

\bibitem[{Abbott {et~al.}(2017{\natexlab{i}})}]{GBM:2017lvd}
---. 2017{\natexlab{i}}, Astrophys. J., 848, L12

\bibitem[{Abbott {et~al.}(2018)}]{LIGOScientific:2018mvr}
---. 2018, arXiv:1811.12907

\bibitem[{Abbott {et~al.}(2019)}]{Abbott:2019yzh}
---. 2019, arXiv:1908.06060

\bibitem[{Acernese {et~al.}(2015)}]{AdV_ref}
Acernese, F., {et~al.} 2015, Class. Quant. Grav., 32, 024001

\bibitem[{Acernese {et~al.}(2018)}]{Acernese:2018bfl}
---. 2018, Class. Quant. Grav., 35, 205004

\bibitem[{Addison {et~al.}(2018)Addison, Watts, Bennett, Halpern, Hinshaw, \&
  Weiland}]{Addison:2017fdm}
Addison, G.~E., Watts, D.~J., Bennett, C.~L., {et~al.} 2018, Astrophys. J.,
  853, 119

\bibitem[{Aghanim {et~al.}(2018)}]{Aghanim:2018eyx}
Aghanim, N., {et~al.} 2018, arXiv:1807.06209

\bibitem[{Ajith(2011)}]{Ajith2011b}
Ajith, P. 2011, Phys.Rev., D84, 084037

\bibitem[{Ajith \& Bose(2009)}]{Ajith:2009fz}
Ajith, P., \& Bose, S. 2009, Phys. Rev., D79, 084032

\bibitem[{Apostolatos {et~al.}(1994)Apostolatos, Cutler, Sussman, \&
  Thorne}]{ACST94}
Apostolatos, T.~A., Cutler, C., Sussman, G.~J., \& Thorne, K.~S. 1994, Phys.
  Rev.~D, 49, 6274

\bibitem[{Arun {et~al.}(2005)Arun, Iyer, Sathyaprakash, \&
  Sundararajan}]{AISS05}
Arun, K.~G., Iyer, B.~R., Sathyaprakash, B.~S., \& Sundararajan, P.~A. 2005,
  Phys.~Rev.~D, 71, 084008, erratum-ibid. ~{\bf D } 72, 069903 (2005)

\bibitem[{Arun {et~al.}(2009)Arun, Mishra, Van Den~Broeck, Iyer, Sathyaprakash,
  \& Sinha}]{AMVISS09}
Arun, K.~G., Mishra, C., Van Den~Broeck, C., {et~al.} 2009, Class. Quant.
  Grav., 26, 094021

\bibitem[{Aso {et~al.}(2013)Aso, Michimura, Somiya, Ando, Miyakawa, Sekiguchi,
  Tatsumi, \& Yamamoto}]{KAGRA_ref}
Aso, Y., Michimura, Y., Somiya, K., {et~al.} 2013, Phys. Rev., D88, 043007

\bibitem[{{Aylor} {et~al.}(2019){Aylor}, {Joy}, {Knox}, {Millea},
  {Raghunathan}, \& {Kimmy Wu}}]{Aylor:2019jkm}
{Aylor}, K., {Joy}, M., {Knox}, L., {et~al.} 2019, \apj, 874, 4

\bibitem[{Bahcall(1995)}]{Bahcall:1995tf}
Bahcall, N.~A. 1995, in {13th Jerusalem Winter School in Theoretical Physics:
  Formation of Structure in the Universe Jerusalem, Israel, 27 December 1995 -
  5 January 1996}

\bibitem[{Betoule {et~al.}(2014)}]{Betoule:2014frx}
Betoule, M., {et~al.} 2014, Astron. Astrophys., 568, A22

\bibitem[{Blair {et~al.}(2008)Blair, Barriga, Brooks, Charlton, Coward, Dumas,
  Fan, Galloway, Gras, Hosken, Howell, Hughes, Ju, McClelland, Melatos, Miao,
  Munch, Scott, Slagmolen, Veitch, Wen, Webb, Wolley, Yan, \& Zhao}]{Blair2008}
Blair, D.~G., Barriga, P., Brooks, A.~F., {et~al.} 2008, Journal of Physics:
  Conference Series, 122, 012001

\bibitem[{{Blanton} {et~al.}(2003){Blanton}, {Hogg}, {Bahcall}, {Brinkmann},
  {Britton}, {Connolly}, {Csabai}, {Fukugita}, {Loveday}, {Meiksin}, {Munn},
  {Nichol}, {Okamura}, {Quinn}, {Schneider}, {Shimasaku}, {Strauss}, {Tegmark},
  {Vogeley}, \& {Weinberg}}]{Blanton:2003hbb}
{Blanton}, M.~R., {Hogg}, D.~W., {Bahcall}, N.~A., {et~al.} 2003, \apj, 592,
  819

\bibitem[{{Bliokh} \& {Minakov}(1975)}]{Bliokh1975}
{Bliokh}, P.~V., \& {Minakov}, A.~A. 1975, \apss, 34, L7

\bibitem[{{Bontz} \& {Haugan}(1981)}]{Bontz1981}
{Bontz}, R.~J., \& {Haugan}, M.~P. 1981, \apss, 78, 199

\bibitem[{Brilenkov {et~al.}(2017)Brilenkov, Eingorn, \&
  Zhuk}]{Brilenkov:2015uxa}
Brilenkov, R., Eingorn, M., \& Zhuk, A. 2017, Astron. Astrophys. Trans., 30, 81

\bibitem[{{Buonanno} {et~al.}(2009){Buonanno}, {Iyer}, {Ochsner}, {Pan}, \&
  {Sathyaprakash}}]{BIOPS2009}
{Buonanno}, A., {Iyer}, B.~R., {Ochsner}, E., {Pan}, Y., \& {Sathyaprakash},
  B.~S. 2009, \prd, 80, 084043

\bibitem[{Cavalier {et~al.}(2006)Cavalier, Barsuglia, Bizouard, Brisson,
  Clapson, Davier, Hello, Kreckelbergh, Leroy, \& Varvella}]{Cavalier:2006rz}
Cavalier, F., Barsuglia, M., Bizouard, M.-A., {et~al.} 2006, Phys. Rev., D74,
  082004

\bibitem[{{Chan} {et~al.}(2018){Chan}, {Messenger}, {Heng}, \&
  {Hendry}}]{Leong2018}
{Chan}, M.~L., {Messenger}, C., {Heng}, I.~S., \& {Hendry}, M. 2018, \prd, 97,
  123014

\bibitem[{Cramer(1946)}]{Cramer46}
Cramer, H. 1946, Mathematical methods in statistics (Princeton University
  Press, NJ, U.S.A.: Pergamon Press)

\bibitem[{Cutler \& Flanagan(1994)}]{CF94}
Cutler, C., \& Flanagan, E. 1994, Phys. Rev. D, 49, 2658

\bibitem[{Dai {et~al.}(2017)Dai, Venumadhav, \& Sigurdson}]{Dai:2016igl}
Dai, L., Venumadhav, T., \& Sigurdson, K. 2017, Phys. Rev., D95, 044011

\bibitem[{Dalal {et~al.}(2006)Dalal, Holz, Hughes, \& Jain}]{Dalal:2006qt}
Dalal, N., Holz, D.~E., Hughes, S.~A., \& Jain, B. 2006, Phys. Rev., D74,
  063006

\bibitem[{{Deguchi} \& {Watson}(1986)}]{Deguchi1986}
{Deguchi}, S., \& {Watson}, W.~D. 1986, \apj, 307, 30

\bibitem[{{Dilday} {et~al.}(2010){Dilday}, {Bassett}, {Becker}, {Bender},
  {Castander}, {Cinabro}, {Frieman}, {Galbany}, {Garnavich}, {Goobar}, {Hopp},
  {Ihara}, {Jha}, {Kessler}, {Lampeitl}, {Marriner}, {Miquel}, {Moll{\'a}},
  {Nichol}, {Nordin}, {Riess}, {Sako}, {Schneider}, {Smith}, {Sollerman},
  {Wheeler}, {{\"O}stman}, {Bizyaev}, {Brewington}, {Malanushenko},
  {Malanushenko}, {Oravetz}, {Pan}, {Simmons}, \& {Snedden}}]{Dilday2010}
{Dilday}, B., {Bassett}, B., {Becker}, A., {et~al.} 2010, \apj, 715, 1021

\bibitem[{Evans {et~al.}(2017)}]{Evans:2017mmy}
Evans, P.~A., {et~al.} 2017, Science, 358, 1565

\bibitem[{Fairhurst(2011)}]{Fairhurst2010}
Fairhurst, S. 2011, Class.Quant.Grav., 28, 105021

\bibitem[{Fishbach {et~al.}(2018)Fishbach, Gray, Hernandez, Qi, \&
  Sur}]{Fishbach:2018gjp}
Fishbach, M., Gray, R., Hernandez, I.~M., Qi, H., \& Sur, A. 2018,
  arXiv:1807.05667

\bibitem[{{Freedman} {et~al.}(2019){Freedman}, {Madore}, {Hatt}, {Hoyt},
  {Jang}, {Beaton}, {Burns}, {Lee}, {Monson}, {Neeley}, {Phillips}, {Rich}, \&
  {Seibert}}]{fmh+19}
{Freedman}, W.~L., {Madore}, B.~F., {Hatt}, D., {et~al.} 2019, arXiv e-prints,
  arXiv:1907.05922

\bibitem[{{Gal-Yam} {et~al.}(2002){Gal-Yam}, {Maoz}, \& {Sharon}}]{gms02}
{Gal-Yam}, A., {Maoz}, D., \& {Sharon}, K. 2002, \mnras, 332, 37

\bibitem[{Gehrels {et~al.}(2016)Gehrels, Cannizzo, Kanner, Kasliwal, Nissanke,
  \& Singer}]{Gehrels:2015uga}
Gehrels, N., Cannizzo, J.~K., Kanner, J., {et~al.} 2016, Astrophys. J., 820,
  136

\bibitem[{{Girardi} {et~al.}(2002){Girardi}, {Manzato}, {Mezzetti}, {Giuricin},
  \& {Limboz}}]{Girardi:2002mmg}
{Girardi}, M., {Manzato}, P., {Mezzetti}, M., {Giuricin}, G., \& {Limboz}, F.
  2002, \apj, 569, 720

\bibitem[{{Graham} {et~al.}(2008){Graham}, {Pritchet}, {Sullivan}, {Gwyn},
  {Neill}, {Hsiao}, {Astier}, {Balam}, {Balland}, {Basa}, {Carlberg}, {Conley},
  {Fouchez}, {Guy}, {Hardin}, {Hook}, {Howell}, {Pain}, {Perrett}, {Regnault},
  {Baumont}, {LeDu}, {Lidman}, {Perlmutter}, {Ripoche}, {Suzuki}, {Walker}, \&
  {Zhang}}]{gps+08}
{Graham}, M.~L., {Pritchet}, C.~J., {Sullivan}, M., {et~al.} 2008, \aj, 135,
  1343

\bibitem[{Gunnarsson {et~al.}(2006)Gunnarsson, Dahlen, Goobar, Jonsson, \&
  Mortsell}]{Gunnarsson:2005qu}
Gunnarsson, C., Dahlen, T., Goobar, A., Jonsson, J., \& Mortsell, E. 2006,
  Astrophys. J., 640, 417

\bibitem[{{Hall} \& {Evans}(2019)}]{Hall2019}
{Hall}, E.~D., \& {Evans}, M. 2019, arXiv e-prints, arXiv:1902.09485

\bibitem[{Hannuksela {et~al.}(2019)Hannuksela, Haris, Ng, Kumar, Mehta, Keitel,
  Li, \& Ajith}]{Hannuksela:2019kle}
Hannuksela, O.~A., Haris, K., Ng, K. K.~Y., {et~al.} 2019, arXiv:1901.02674

\bibitem[{Hearin {et~al.}(2017)}]{Hearin:2016uxs}
Hearin, A.~P., {et~al.} 2017, Astron. J., 154, 190

\bibitem[{Hotokezaka {et~al.}(2019)Hotokezaka, Nakar, Gottlieb, Nissanke,
  Masuda, Hallinan, Mooley, \& Deller}]{Hotokezaka:2018dfi}
Hotokezaka, K., Nakar, E., Gottlieb, O., {et~al.} 2019, Nature Astron.,
  arXiv:1806.10596

\bibitem[{Humphreys {et~al.}(2013)Humphreys, Reid, Moran, Greenhill, \&
  Argon}]{Humphreys:2013eja}
Humphreys, E. M.~L., Reid, M.~J., Moran, J.~M., Greenhill, L.~J., \& Argon,
  A.~L. 2013, Astrophys. J., 775, 13

\bibitem[{Iyer {et~al.}(2011)}]{Ligo-india}
Iyer, B., {et~al.} 2011, LIGO-India Technical Report No. LIGO-M1100296

\bibitem[{{Jones} {et~al.}(2018){Jones}, {Riess}, {Scolnic}, {Pan}, {Johnson},
  {Coulter}, {Dettman}, {Foley}, {Foley}, {Huber}, {Jha}, {Kilpatrick},
  {Kirshner}, {Rest}, {Schultz}, \& {Siebert}}]{jrs+18}
{Jones}, D.~O., {Riess}, A.~G., {Scolnic}, D.~M., {et~al.} 2018, \apj, 867, 108

\bibitem[{Karki {et~al.}(2016)}]{Karki:2016pht}
Karki, S., {et~al.} 2016, Rev. Sci. Instrum., 87, 114503

\bibitem[{Kastha {et~al.}(2018)Kastha, Gupta, Arun, Sathyaprakash, \& Van
  Den~Broeck}]{Kastha2018}
Kastha, S., Gupta, A., Arun, K.~G., Sathyaprakash, B.~S., \& Van Den~Broeck, C.
  2018, Phys. Rev., D98, 124033

\bibitem[{Keeley {et~al.}(2019)Keeley, Shafieloo, L'Huillier, \&
  Linder}]{Keeley:2019hmw}
Keeley, R.~E., Shafieloo, A., L'Huillier, B., \& Linder, E.~V. 2019,
  arXiv:1905.10216

\bibitem[{Kocsis {et~al.}(2006)Kocsis, Frei, Haiman, \& Menou}]{KFHM06}
Kocsis, B., Frei, Z., Haiman, Z., \& Menou, K. 2006, Astrophys. J., 637, 27

\bibitem[{Krolak \& Schutz(1987)}]{Krolak1987}
Krolak, A., \& Schutz, B.~F. 1987, General Relativity and Gravitation, 19, 1163

\bibitem[{{Li} {et~al.}(2011){Li}, {Chornock}, {Leaman}, {Filippenko},
  {Poznanski}, {Wang}, {Ganeshalingam}, \& {Mannucci}}]{2011MNRAS.412.1473L}
{Li}, W., {Chornock}, R., {Leaman}, J., {et~al.} 2011, \mnras, 412, 1473

\bibitem[{Lokas \& Mamon(2003)}]{Lokas:2003ks}
Lokas, E.~L., \& Mamon, G.~A. 2003, Mon. Not. Roy. Astron. Soc., 343, 401

\bibitem[{{LSST Science Collaboration} {et~al.}(2017){LSST Science
  Collaboration}, {Marshall}, {Anguita}, {Bianco}, {Bellm}, {Brandt},
  {Clarkson}, {Connolly}, {Gawiser}, {Ivezic}, {Jones}, {Lochner}, {Lund},
  {Mahabal}, {Nidever}, {Olsen}, {Ridgway}, {Rhodes}, {Shemmer}, {Trilling},
  {Vivas}, {Walkowicz}, {Willman}, {Yoachim}, {Anderson}, {Antilogus}, {Angus},
  {Arcavi}, {Awan}, {Biswas}, {Bell}, {Bennett}, {Britt}, {Buzasi},
  {Casetti-Dinescu}, {Chomiuk}, {Claver}, {Cook}, {Davenport}, {Debattista},
  {Digel}, {Doctor}, {Firth}, {Foley}, {Fong}, {Galbany}, {Giampapa}, {Gizis},
  {Graham}, {Grillmair}, {Gris}, {Haiman}, {Hartigan}, {Hawley}, {Hlozek},
  {Jha}, {Johns-Krull}, {Kanbur}, {Kalogera}, {Kashyap}, {Kasliwal}, {Kessler},
  {Kim}, {Kurczynski}, {Lahav}, {Liu}, {Malz}, {Margutti}, {Matheson},
  {McEwen}, {McGehee}, {Meibom}, {Meyers}, {Monet}, {Neilsen}, {Newman},
  {O'Dowd}, {Peiris}, {Penny}, {Peters}, {Poleski}, {Ponder}, {Richards},
  {Rho}, {Rubin}, {Schmidt}, {Schuhmann}, {Shporer}, {Slater}, {Smith},
  {Soares-Santos}, {Stassun}, {Strader}, {Strauss}, {Street}, {Stubbs},
  {Sullivan}, {Szkody}, {Trimble}, {Tyson}, {de Val-Borro}, {Valenti},
  {Wagoner}, {Wood-Vasey}, \& {Zauderer}}]{lsst_2017}
{LSST Science Collaboration}, {Marshall}, P., {Anguita}, T., {et~al.} 2017,
  arXiv e-prints, arXiv:1708.04058

\bibitem[{Macaulay {et~al.}(2018)}]{Macaulay:2018fxi}
Macaulay, E., {et~al.} 2018, Submitted to: Mon. Not. Roy. Astron. Soc.,
  arXiv:1811.02376

\bibitem[{{Mannucci} {et~al.}(2008){Mannucci}, {Maoz}, {Sharon}, {Botticella},
  {Della Valle}, {Gal-Yam}, \& {Panagia}}]{mms+08}
{Mannucci}, F., {Maoz}, D., {Sharon}, K., {et~al.} 2008, \mnras, 383, 1121

\bibitem[{Messenger \& Read(2012)}]{read}
Messenger, C., \& Read, J. 2012, Phys. Rev. Lett., 108, 091101

\bibitem[{Messenger {et~al.}(2014)Messenger, Takami, Gossan, Rezzolla, \&
  Sathyaprakash}]{Messenger:2013fya}
Messenger, C., Takami, K., Gossan, S., Rezzolla, L., \& Sathyaprakash, B.~S.
  2014, Phys. Rev., X4, 041004

\bibitem[{{Mooley} {et~al.}(2018){Mooley}, {Deller}, {Gottlieb}, {Nakar},
  {Hallinan}, {Bourke}, {Frail}, {Horesh}, {Corsi}, \&
  {Hotokezaka}}]{2018Natur.561..355M}
{Mooley}, K.~P., {Deller}, A.~T., {Gottlieb}, O., {et~al.} 2018, \nat, 561, 355

\bibitem[{Nair {et~al.}(2018)Nair, Bose, \& Saini}]{Nair:2018ign}
Nair, R., Bose, S., \& Saini, T.~D. 2018, Phys. Rev., D98, 023502

\bibitem[{Nakamura(1998)}]{Nakamura1998}
Nakamura, T.~T. 1998, Phys. Rev. Lett., 80, 1138

\bibitem[{Navarro {et~al.}(1996)Navarro, Frenk, \& White}]{Navarro:1995iw}
Navarro, J.~F., Frenk, C.~S., \& White, S. D.~M. 1996, Astrophys. J., 462, 563

\bibitem[{Nissanke {et~al.}(2010)Nissanke, Holz, Hughes, Dalal, \&
  Sievers}]{Nissanke:2009kt}
Nissanke, S., Holz, D.~E., Hughes, S.~A., Dalal, N., \& Sievers, J.~L. 2010,
  Astrophys. J., 725, 496

\bibitem[{Ohanian(1974)}]{Ohanian1974}
Ohanian, H.~C. 1974, International Journal of Theoretical Physics, 9, 425

\bibitem[{{Phillips} {et~al.}(1999){Phillips}, {Lira}, {Suntzeff}, {Schommer},
  {Hamuy}, \& {Maza}}]{pls+99}
{Phillips}, M.~M., {Lira}, P., {Suntzeff}, N.~B., {et~al.} 1999, \aj, 118, 1766

\bibitem[{Pietrzyński {et~al.}(2013)}]{Pietrzynski:2013gia}
Pietrzyński, G., {et~al.} 2013, Nature, 495, 76

\bibitem[{Punturo {et~al.}(2010)}]{ET_ref}
Punturo, M., {et~al.} 2010, Classical and Quantum Gravity, 27, 084007

\bibitem[{Rao(1945)}]{Rao45}
Rao, C. 1945, Bullet. Calcutta Math. Soc, 37, 81

\bibitem[{{Raskin} {et~al.}(2012){Raskin}, {Scannapieco}, {Fryer},
  {Rockefeller}, \& {Timmes}}]{Raskin2012}
{Raskin}, C., {Scannapieco}, E., {Fryer}, C., {Rockefeller}, G., \& {Timmes},
  F.~X. 2012, \apj, 746, 62

\bibitem[{Riess {et~al.}(2019)Riess, Casertano, Yuan, Macri, \&
  Scolnic}]{Riess:2019cxk}
Riess, A.~G., Casertano, S., Yuan, W., Macri, L.~M., \& Scolnic, D. 2019,
  arXiv:1903.07603

\bibitem[{{Riess} {et~al.}(1996){Riess}, {Press}, \& {Kirshner}}]{rpk96}
{Riess}, A.~G., {Press}, W.~H., \& {Kirshner}, R.~P. 1996, \apj, 473, 88

\bibitem[{Riess {et~al.}(2016)}]{Riess:2016jrr}
Riess, A.~G., {et~al.} 2016, Astrophys. J., 826, 56

\bibitem[{Riess {et~al.}(2018)}]{Riess:2018byc}
---. 2018, Astrophys. J., 861, 126

\bibitem[{Rodney {et~al.}(2015)}]{Rodney:2015cwa}
Rodney, S.~A., {et~al.} 2015, Astron. J., 150, 156, [Astron. J.151,47(2016)]

\bibitem[{Sathyaprakash \& Schutz(2009)}]{SathyaSchutzLivRev09}
Sathyaprakash, B., \& Schutz, B. 2009, Living Rev.Rel., 12, 2

\bibitem[{Schutz(1986)}]{Schutz86}
Schutz, B.~F. 1986, Nature (London), 323, 310

\bibitem[{Scolnic {et~al.}(2018)}]{Scolnic:2017caz}
Scolnic, D.~M., {et~al.} 2018, Astrophys. J., 859, 101

\bibitem[{{Shapiro} {et~al.}(2010){Shapiro}, {Bacon}, {Hendry}, \&
  {Hoyle}}]{Shapiro2010}
{Shapiro}, C., {Bacon}, D.~J., {Hendry}, M., \& {Hoyle}, B. 2010, \mnras, 404,
  858

\bibitem[{{Sharon} {et~al.}(2007){Sharon}, {Gal-Yam}, {Maoz}, {Filippenko}, \&
  {Guhathakurta}}]{sgm+07}
{Sharon}, K., {Gal-Yam}, A., {Maoz}, D., {Filippenko}, A.~V., \&
  {Guhathakurta}, P. 2007, \apj, 660, 1165

\bibitem[{Somiya(2012)}]{Somiya:2011np}
Somiya, K. 2012, Class. Quant. Grav., 29, 124007

\bibitem[{Tagoshi {et~al.}(2014)Tagoshi, Mishra, Pai, \& Arun}]{TMPA2014}
Tagoshi, H., Mishra, C.~K., Pai, A., \& Arun, K. 2014, Phys.Rev., D90, 024053

\bibitem[{Tuyenbayev {et~al.}(2017)}]{Tuyenbayev:2016xey}
Tuyenbayev, D., {et~al.} 2017, Class. Quant. Grav., 34, 015002

\bibitem[{Usman {et~al.}(2018)Usman, Mills, \& Fairhurst}]{Usman:2018imj}
Usman, S.~A., Mills, J.~C., \& Fairhurst, S. 2018, arXiv:1809.10727

\bibitem[{Van Den~Broeck {et~al.}(2010)Van Den~Broeck, Trias, Sathyaprakash, \&
  Sintes}]{VanDenBroeck:2010fp}
Van Den~Broeck, C., Trias, M., Sathyaprakash, B.~S., \& Sintes, A.~M. 2010,
  Phys. Rev., D81, 124031

\bibitem[{Viets {et~al.}(2018)}]{Viets:2017yvy}
Viets, A., {et~al.} 2018, Class. Quant. Grav., 35, 095015

\bibitem[{Vitale \& Chen(2018)}]{Vitale:2018wlg}
Vitale, S., \& Chen, H.-Y. 2018, Phys. Rev. Lett., 121, 021303

\bibitem[{Wang {et~al.}(1996)Wang, Stebbins, \& Turner}]{Wang:1996as}
Wang, Y., Stebbins, A., \& Turner, E.~L. 1996, Phys. Rev. Lett., 77, 2875

\bibitem[{Wen \& Chen(2010)}]{Wen:2010cr}
Wen, L., \& Chen, Y. 2010, Phys. Rev., D81, 082001

\bibitem[{{Wong} {et~al.}(2019){Wong}, {Suyu}, {Chen}, {Rusu}, {Millon},
  {Sluse}, {Bonvin}, {Fassnacht}, {Taubenberger}, {Auger}, {Birrer}, {Chan},
  {Courbin}, {Hilbert}, {Tihhonova}, {Treu}, {Agnello}, {Ding}, {Jee},
  {Komatsu}, {Shajib}, {Sonnenfeld}, {Bland ford}, {Koopmans}, {Marshall}, \&
  {Meylan}}]{2019arXiv190704869W}
{Wong}, K.~C., {Suyu}, S.~H., {Chen}, G. C.~F., {et~al.} 2019, arXiv e-prints,
  arXiv:1907.04869

\bibitem[{Zhao \& Santos(2017)}]{Zhao:2017imr}
Zhao, W., \& Santos, L. 2017, arXiv:1710.10055

\bibitem[{{Zhao} \& {Wen}(2018)}]{Wen2018}
{Zhao}, W., \& {Wen}, L. 2018, \prd, 97, 064031

\end{thebibliography}
\end{document}